\documentclass[11pt,a4paper,tightenlines,eqsecnum,floats,aps,amsmath,amssymb,nofootinbib,prd,showpacs]{revtex4-1}
\pdfoutput=1

\usepackage{epsfig}
\usepackage{setspace}
\usepackage{amsmath,amssymb,amsfonts,amsthm}
\usepackage{graphicx}
\usepackage{multirow}
\usepackage{xfrac}
\usepackage[justification=centering]{caption}
\usepackage{subcaption}
\usepackage{hyperref}
\usepackage{changepage}

\newcommand{\be}{\begin{equation}}
\newcommand{\ee}{\end{equation}}
\newcommand{\beq}{\begin{eqnarray}}
\newcommand{\eeq}{\end{eqnarray}}

\def\lsim{\hbox{ \raise.35ex\rlap{$<$}\lower.6ex\hbox{$\sim$}\ }}
\def\gsim{\hbox{ \raise.35ex\rlap{$>$}\lower.6ex\hbox{$\sim$}\ }} 

\begin{document}
\begin{flushleft}
KCL-PH-TH/2014-7
\end{flushleft}

\title{Cusps and pseudocusps in strings with Y-junctions}

\author{
Thomas Elghozi$^1$\email{thomas.elghozi@kcl.ac.uk} 
William Nelson$^2$\email{wnt20@hotmail.co.uk},
Mairi Sakellariadou$^1$\email{mairi.sakellariadou@kcl.ac.uk}
}

\affiliation{$^1$Department of Physics, King's College London,
University of London,\\Strand WC2R 2LS, London, U.K.\\
$^2$Institute for Mathematics, Astrophysics and Particle
Physics, Radboud University,\\ Heyendaalsweg 135, 6525-AJ Nijmegen,
Netherlands}

\begin{abstract}
We study the occurrence of cuspy events on a light string stretched between two Y-junctions with fixed heavy strings. We first present an analytic study and give a solid criterion to discriminate between cuspy and noncuspy string configurations. We then describe a numerical code, built to test this analysis. Our numerical investigation allows us to look at the correlations between the string network's parameters and the occurrence of cuspy phenomena. We show that the presence of large-amplitude waves on the light string leads to cuspy events. We then relate the occurrence of cuspy events to features like the number of vibration modes on the string or the string's root-mean-square velocity.
\end{abstract}

\keywords{cosmic strings, gravitational waves, cosmological applications of theories with extra dimensions}

\maketitle

\newpage
\section{Introduction}

Cosmic strings~\cite{Kibble, Vilenkin_shellard, Hind_Kibble, ms-cs07} can arise as a result of phase transitions followed by spontaneously symmetry breakings in the early Universe. Such one-dimensional false vacuum remnants were shown~\cite{Jeannerot:2003qv,ms-cs08} to be generically formed at the end of hybrid inflation within the context of grand unified theories. The evolution of a cosmic string network has been the core of many analytical and numerical studies. It has been long known and well-accepted that long strings enter the {\sl scaling regime}, rendering a cosmic string network cosmologically acceptable. Much later it was also shown~\cite{Ringeval:2005kr}, by means of numerical simulations, that cosmic string loops in an expanding universe also achieve a scaling solution, and an analytical model has been proposed~\cite{Lorenz:2010sm} to derive the expected number density distribution of cosmic string loops at any redshift soon after the time of string formation to today.

Cosmic superstrings~\cite{PolchRevis,Sakellariadou:2008ie}, the string theory analogues of the solitonic strings, are generically formed \cite{Sarangi:2002yt} at the end of brane inflation. In contrast to the abelian field theory strings which can only interact through intercommutation and exchange of partners with probability of order unity, collisions of cosmic superstrings typically occur with smaller than unity probabilities and can lead to the formation of Y-junctions at which three strings meet~\cite{PolchProb,cmp,JJP}. This characteristic property of cosmic superstrings is of particular interest since it can strongly effect the dynamics of the network evolution~\cite{Sakellariadou:2004wq,TWW,NAVOS,Davis:2008kg,PACPS} leading to potentially observable phenomenological signatures~\cite{PolchRevis,Davis:2008kg,CPRev,ACMPPS}.

The effect of junctions on the evolution of cosmic superstring networks was the central subject of several numerical~\cite{Rajantie:2007hp,Urrestilla:2007yw,Sakellariadou:2008ay,Bevis:2009az} and analytical~\cite{Sakellariadou:2004wq,TWW,NAVOS,Copeland:2006eh,Davis:2008kg,Copeland:2006if,Copeland:2007nv,ACMPPS,PACPS} studies.

One of the most important channels of radiaton emission from cosmic (super)strings is gravity waves~\cite{Vachaspati:1984gt,Sakellariadou:1990ne,Damour:2000wa,Damour:2001bk,Brandenberger:2008ni,Abbott:2009rr,Olmez:2010bi,Binetruy:2010cc,Regimbau:2011bm}. They can be emitted either as bursts, namely by cusps and kinks, or as a stochastic background. To estimate the emission of gravity waves from cosmic (super)strings it is therefore crucial to evaluate the influence of some parameters, such as the interstring distance, the coherence distance and the wiggliness, on the number of cusps. It is usually assumed that cusps appear on the string and their number is just considered as a free and unknown parameter, to be estimated, for example, from numerical simulations. The aim of this analysis is to roughly evaluate the occurrence of cusps on a string network and in particular to relate the probability of cusp's formation to the relevant string parameters.

In what follows, we present first an analytical and then a numerical study of a string stretched between two junctions, and its periodic non-interacting evolution. We consider the specific confi\-gu\-ration of two equal tension heavy strings linked by a light string. As explained in the following, the conclusions drawn in such case can be generalised to realistic strings configurations under certain circumstances which we discuss in Section~\ref{sec:setup}. We estimate the influence of the string parameters on the average number of cuspsy events appearing on the string during its evolution. In particular, we first look at the periodicity requirements and symmetries on the string, in order to allow for a Fourier decomposition. An analytical study then draws a link between waves and cuspy phenomena on the string, where by cuspy phenomena we mean both cusps and \emph{pseudocusps}. Recall that the former are points on the string reaching temporarily the speed of light $c = 1$. The latter are highly relativistic configurations close to cusps but reaching a velocity between $10^{-3}$ and $10^{-6}$ below $c$. We then present our numerical simulation which allows us to draw a specific string configuration and to subsequently compute the number of cusps and pseudocusps within a period of a non-interacting evolution. Finally, we discuss our results with respect to two parameters, one that sets the interstring distance and another one that measures the waviness of the string --- that is how many large-amplitude waves are on the string and how large there are.

\section{General setup} \label{sec:setup}

In the context of string theory, stable bound states of fundamental strings and one-dimensional Dirichlet branes can be formed, leading to the emergence of Y-junctions~\cite{Copeland:2003bj}. These junctions can also appear in the context of semi-local string interactions. These types of strings are thought to have generically cusps, especially in the case of a string stretched between two junctions~\cite{Davis:2008kg}. Here, we start with a simplified and idealised version of such a configuration in order to look at the parameters influencing the occurrence and number of cusps.

The Y-junction configuration we will study is made our of two heavy strings connected via a light string. Hence, without loss of generality we consider the heavy strings to be of equal tension.\footnote{The formation of a junction depends on various parameters, such as the collision velocity and the tensions. However, once the junction is formed, the tensions will not influence the dynamics~\cite{Copeland:2006if}.} So in what follows, we have two tensions: the tension of the heavy strings and that of the light string.

\begin{figure}[t]
	\begin{center}
	\includegraphics[height=6.38cm, width=10cm]{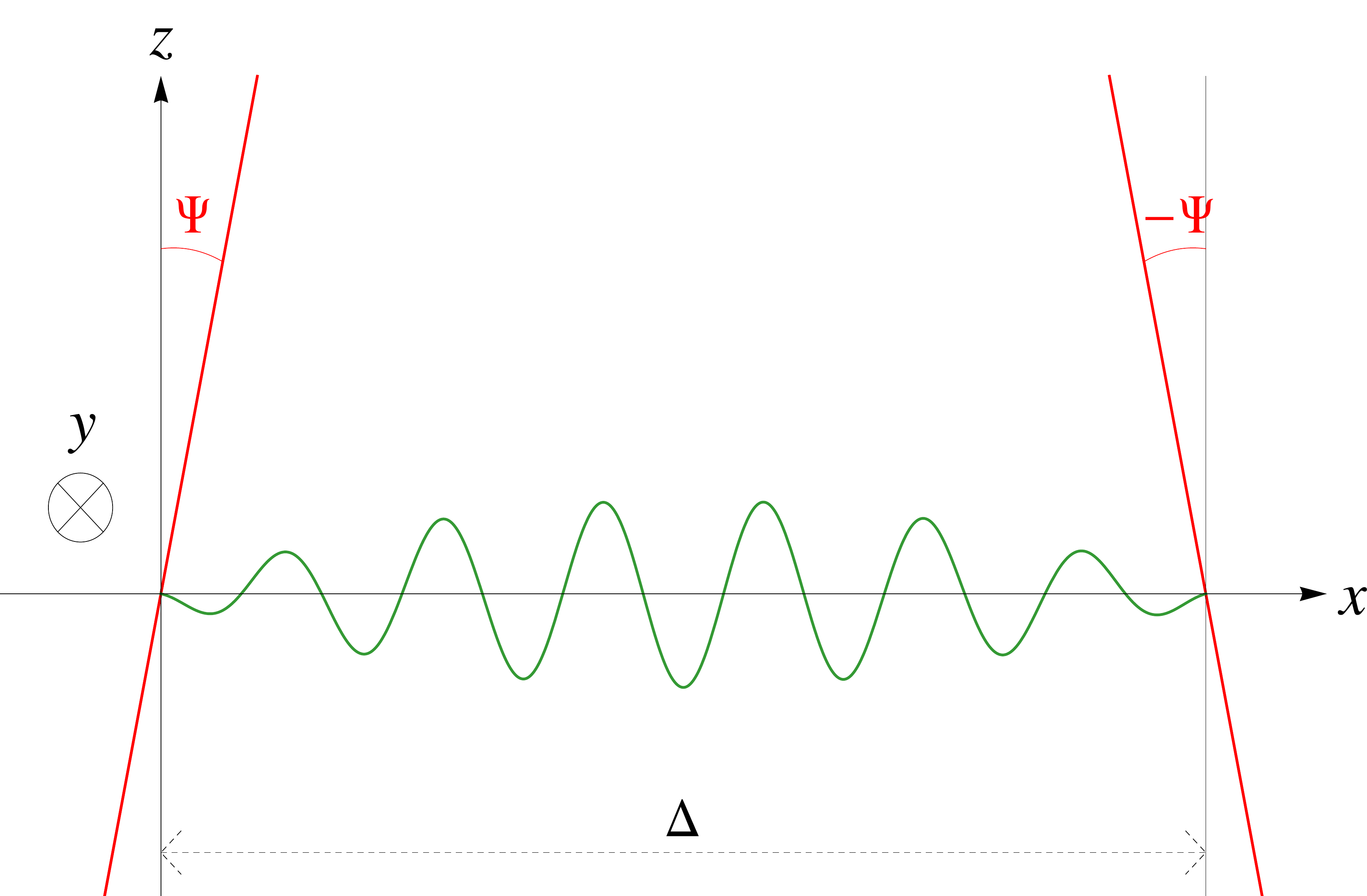}
	\caption{A light string stretched between two junctions with heavy strings.}
	\label{StringLayout}
	\end{center}
\vspace{-1em}\end{figure}

We here consider two heavy strings in the $xz$-plane, oriented along the $z$-axis, and then we tilt them by an angle $\pm \Psi$ with respect to the $z$-direction (see Fig.~\ref{StringLayout}) and space them out by a distance $\Delta$. The heavy strings are considered heavy enough to be at rest at least for a time longer than the time scale of the light string's movement. This implies either that the heavy string's tension is very large compared to (at least of order $10^2$ times) the light string's one, or at least that the time scale of the light string's movement is short compared to the ratio of the light string's length to the heavy string's velocity (with respect to the light string). In addition, since the heavy strings can be considered as straight in the vicinity of the junction and since the boundary conditions are what matter here, the heavy strings will be taken infinitely straight. Note that even though the case studied here is not generic, the conclusions are applicable to generalisations of this specific configuration as shown in the appendix.

The boundary conditions for a light string ending on two junctions with the aforementioned heavy strings are given by
\begin{subequations} \label{condattheborder}
	\begin{align}
	\dot{\bf x}_\perp \left( t,0 \right) &= {\bf x}'_\parallel\left( t,0 \right) = 0~, \\
	\dot{\bf x}_\perp \left( t,\sigma_{m} \right) &= {\bf x}'_\parallel\left( t,\sigma_{m} \right) = 0~,
	\end{align}
\end{subequations}
where $f'(\sigma,t) = \partial_\sigma f(\sigma,t)$ and $\dot f(\sigma,t) = \partial_t f(\sigma,t)$ and where the subscripts $\perp$, $\parallel$ indicate the projection along the directions orthogonal, parallel to the (local) end string, respectively. The string's position vector ${\bf x}$ depends on two world-sheet coordinates, namely the cosmological time $t$\footnote{One can indeed choose to work in the time gauge, so that the time-like coordinate $\tau$ is indeed the cosmological time $t$.} and the space-like coordinate $\sigma \in [\:\! 0,\sigma_{m}]$, denoting the position on the string, with $\sigma_{m}$ being the parameter length of the string, that is the maximal value for $\sigma$ since the minimal value is $0$. Hence, in terms of the coordinates ($x, y, z$) of ${\bf x}$, conditions (\ref{condattheborder}), at any time $t$, read
\begin{subequations}
	\begin{align}
	\dot{x}_y \left( t,0\right) &= 0~, \\
	\dot{x}_x \left( t,0\right) \cos \Psi - \dot{x}_z\left( t,0\right) \sin \Psi &=0~, \\
	x'_x \left( t,0\right) \sin \Psi + x'_z\left( t,0\right) \cos \Psi &= 0~,
	\end{align}
and\vspace{-1.2em}
	\begin{align}
	\dot{x}_y \left( t,\sigma_{m}\right) &= 0~, \\
	\dot{x}_x \left( t,\sigma_{m}\right) \cos \Psi + \dot{x}_z\left( t,\sigma_{m}\right) \sin \Psi &= 0~, \\
	- x'_x \left( t,\sigma_{m}\right) \sin \Psi + x'_z\left( t,\sigma_{m}\right) \cos \Psi &= 0~.
	\end{align}
\end{subequations}
Following the usual approach, one imposes the conformal gauge conditions $(\dot{x}_\mu)^2 + (x_\mu')^2= 0$ and $\dot{x}^\mu \,x_\mu' = 0$ and the temporal gauge $\tau = t \equiv x^0$, to get $\bf{x}'' - {\bf\ddot{x}} = {\bf 0}$. To solve this equation we decompose the position vectors into left- and right-movers, ${\bf a} \left( \sigma+t \right)$, ${\bf b} \left( \sigma-t \right)$, as
\be
	{\bf x} \left( t , \sigma \right) \equiv \frac{1}{2} \left[ {\bf a} \left( \sigma+t \right) + {\bf b} \left( \sigma-t \right) \right]~,
\ee
leading to the system of equations
\begin{subequations} \label{AandBx6}
	\begin{align}
	a_y' \left( t \right) &= b_y' \left( -t \right)~, \label{eq:1}\\
	\left[\, a_z' \left( t \right) - b_z' \left( -t \right) \:\!\right]\, \tan \Psi &= a_x' \left( t \right) - b_x' \left( -t \right)~, \label{eq:2}\\
	a_z' \left( t \right) + b_z' \left( -t \right) &= - \left[\, a_x' \left( t \right) + b_x' \left( -t \right) \:\!\right]\, \tan \Psi~, \label{eq:3}
	\end{align}
and\vspace{-1.2em}
	\begin{align}
	a_y' \left( \sigma_{m}+t \right) &= b_y' \left( \sigma_{m}-t \right)~, \label{eq:4}\\
	\left[\, a_z' \left( \sigma_{m}+t \right) - b_z' \left( \sigma_{m}-t \right) \:\!\right]\,\tan \Psi &= - a_x' \left( \sigma_{m}+t \right) + b_x' \left( \sigma_{m}-t \right)~, \label{eq:5}\\
	a_z' \left( \sigma_{m}+t \right) + b_z' \left( \sigma_{m}-t \right) &= \left[\, a_x' \left( \sigma_{m}+t \right) + b_x' \left( \sigma_{m}-t \right) \:\!\right]\, \tan \Psi~. \label{eq:6}
	\end{align}
\end{subequations}

\subsection{Periodicity requirements}

Equations~(\ref{eq:1}) and (\ref{eq:4}) imply
\be \label{eq:periodz}
	a_y' \left( -\sigma_{m}+t \right) = a_y' \left( \sigma_{m}+t \right)~,
\ee
namely that $a_y' \left( \sigma+t \right)$ (and hence $b_y' \left( \sigma-t \right)$) is $2 \sigma_m$-periodic.

Redefining $t \rightarrow t + \sigma_{m}$ in Eqs.~(\ref{eq:5}) and~(\ref{eq:6}) and combining with Eqs.~(\ref{eq:2}) and~(\ref{eq:3}), we get the difference equation:
\beq \label{eq:diff1}
	a_z' \left( t \right) = - {\cal R} a_z' \left( -2\sigma_{m}+t \right) - a_z' \left( -4\sigma_{m}+t \right)~,
\eeq
where
\be
	{\cal R} \equiv -2 \cos ( 4 \Psi )~,
\ee
and similarly for $a_x' \left( t \right)$. Setting $t\rightarrow t - 2 n \sigma_{m}$ and defining
\be
	a_n \equiv a_x' \left( -2 \left( n+1\right) \sigma_{m} + t \right) \quad \left( \mbox{or similarly }\; a_z' \left( -2 \left( n+1\right) \sigma_{m} + t \right) \right)~,
\ee
Eq.~(\ref{eq:diff1}) reads
\be
	a_{n+2} = - {\cal R} a_{n+1} - a_{n}~.
\ee
with general solution
\be \label{eq:general_sol}
	a_n = 2 E \cos \left( n \bar{\Psi} \right) + 2 F \sin \left( n \bar{\Psi} \right)~,
\ee
where $\bar{\Psi} = \arccos \left( -{\cal R}/2 \right) = 4 \Psi \;\mathrm{mod}\, 2 \pi$, and the constants $E$ and $F$ are chosen to give $a_0$ and $a_1$ (i.e., $a_x'\left( -2\sigma_{m}+t \right)$ and $a_x'\left( -4\sigma_{m}+t \right)$).

We want to determine if the function $a_n$ is periodic, i.e. we want to find $m\in {\mathbb Z}$ so that $a_m = a_0$. Note that in such case, $a_n$ is $m$-periodic and $a'_x$ and $a'_z$ are $2 m \sigma_m$-periodic. From Eq.~(\ref{eq:general_sol}), it is clear that this occurs for
\be
	m = \frac{2 \pi M}{\bar{\Psi}}~,
\ee
where $M\in {\mathbb Z}$. Using the definition of $\bar{\Psi}$, we find that such a solution exists provided
\be \label{eq:cond1}
	\bar{\Psi} = \arccos \left( -{\cal R}/2 \right) = \frac{p \pi}{q} \quad \Leftrightarrow \quad 4 \Psi = \frac{p \pi}{q}~,
\ee
for $p,q \in {\mathbb Z}$, for which the function $a_n$ is then periodic in $a_0 = a_m = a_{2 q M /p}$, for any arbitrary integer $M$; that is the function $a_x' \left( \sigma+t \right)$ is then periodic in $\sigma \rightarrow \sigma + 2qM\sigma_{m} /p$, for any arbitrary integer $M$.

Solving Eq.~(\ref{eq:cond1}) for $\Psi$, we find that the function $a'_x\left( \sigma+t \right)$ is periodic with period $2 \sigma_m /Q$ provided
\be
	\Psi = \frac{1}{2} \arctan \left( \pm \sqrt{ \frac{1 - \cos \left( Q \pi\right) }{1 + \cos \left( Q \pi\right) } }\right) \quad \Leftrightarrow \quad \Psi = \frac{Q \pi}{4}~,
\ee
where $Q$ is a rational: $Q \in {\mathbb Q}$. Thus, for a dense subset of angles in the range $\Psi \in [ -\pi/2 ,\, \pi/2 ]$, $a_x' \left( \sigma+t \right)$ and $a_z'\left( \sigma+t \right)$ are periodic, and hence they can be decomposed in a Fourier series to simplify the analysis.

Concern over what happens for angles not satisfying Eq.~(\ref{eq:cond1}) can be alleviated by noting that although the functions $a'_x\left( \sigma+t \right)$ are not periodic, they are arbitrarily close to periodic, and this is sufficient for our requirements here, that is for our qualitative study. It might also be worth noting that the period can be large, which might cause problems for our approximation namely that the end strings are static over one period --- indeed, if the period is very long, the heavy strings cannot be considered static over such a large time scale anymore.

Finally, recall this specific setup is considered for its simplicity. The conclusions on the overall periodicity or quasi-periodicity, drawn from the above analysis, are thought to be generic though, since the configuration choices made here leave the string's dynamical properties unchanged. In addition, we studied in the appendix how these results on periodicity are modified in a more realistic and more complex strings configuration, confirming our initial intuition.

\subsection{Symmetries}

To proceed, let us focus on the symmetries between the two movers on the string. Using Eqs.~(\ref{AandBx6}), we obtain
\begin{subequations}
	\begin{align}
	b_x' \left( -t \right) &= \frac{1}{1+\tan^2\Psi} \left( \left( 1-\tan^2\Psi \right) a_x' \left( t \right) - 2 \tan\Psi \;a_z' \left( t \right) \right)~, \\
	b_z' \left( -t \right) &= \frac{-1}{1+\tan^2\Psi} \left( \left( 1-\tan^2\Psi \right) a_z' \left( t \right) + 2 \tan\Psi \; a_x' \left( t \right) \right)~, \\
	b_y' \left( -t \right) &= a_y' \left( t \right)~.
	\end{align}
\end{subequations}
Since ${\bf b}'( \sigma-t ) = {\bf x}' (\sigma,t) - \dot{\bf x} (\sigma,t)$ we remark that ${\bf b}'(-t) = - {\bf b}'(t)$, and then writing the above set of equations in vector notation, we get
\be
	{\bf b}' \left( t \right) = {\bf T} \, {\bf a}' \left( t \right)~,
\ee
where the matrix ${\bf T}$ is defined by
\be
	{\bf T} = \left( \begin{array}{ccc}
		-\frac{1-\tan^2\Psi}{1+\tan^2\Psi} & 0 & \frac{2\tan \Psi}{1+\tan^2\Psi} \\
		0 & -1 & 0 \\
		\frac{2\tan \Psi}{1+\tan^2\Psi} & 0 & \frac{1-\tan^2\Psi}{1+\tan^2\Psi}
	\end{array} \right) = \left(\begin{array}{ccc}
		-\cos\left( 2\Psi \right) & 0 & \sin\left( 2\Psi \right) \\
		0 & -1 & 0 \\
		\sin\left( 2\Psi \right) & 0 & \cos\left( 2\Psi \right)
	\end{array} \right)~.
\ee
This matrix is diagonalised by a change of basis, such that the $z$-axis is parallel to the $\sigma=0$ end string. In this basis, we get
\be \label{eq:sym}
	{\bf b}'\left( t \right) = \left( \begin{array}{ccc}
		\!-1 & \,0 & \;\,0\, \\
		0 & -1 & \;\,0\, \\
		0 & \,0 & \;\,1\,
	\end{array} \right) {\bf a}'\left( t \right)~.
\ee
Thus, ${\bf b}'\left(t\right)$ is simply given by a reflection of ${\bf a}'\left(t\right)$ through the axis parallel to the end string.

Note in particular that the square velocity of the string is
\beq \label{eq:vel}
	{\bf v}.{\bf v}\left( t,\sigma \right) &=& \frac{1}{2} \left( 1 + {\bf a}' \left( \sigma+t \right).{\bf b}' \left( \sigma-t \right) \right) \\
	&=& \frac{1}{2} \left( 1 + a_\parallel' \left( \sigma+t \right) a_\parallel' \left( \sigma+t \right) - {\bf a}_\perp' \left( \sigma+t \right) {\bf a}_\perp' \left( \sigma+t \right) \right)~,
\eeq
where $a'_\parallel$ and $a'_\perp$ are the components of ${\bf a}'$ parallel and perpendicular to the ($\sigma =0$) end string, respectively.

\section{The probability of cusps and pseudocusps}

Let us recall that cusps appear when the two curves ${\bf a'}$ and ${\bf -b'}$ cross each other on the unit sphere --- remembering that $|{\bf a'}| = 1 = |{\bf b'}|$ as a consequence of the Visaroso condition. This is equivalent to defining cusps as points reaching, for some instant $t$, the speed of light $c = 1$. Indeed, $\dot {\bf x} (\sigma,t) \equiv \sfrac{1}{2} \, ({\bf a'}(\sigma+t) - {\bf b'} (\sigma-t)) = {\bf a'} (\sigma+t) = {\bf b'} (\sigma-t)$ in the case of cusps.

There is a similar event we will address, and we will refer to as a \emph{pseudo-cusp}, which occurs when the two curves ${\bf a'}$ and ${\bf -b'}$ are very close (and we will see how close) to each other, without however intersecting. Pseudocusps have to be considered firstly because when trying to determine statistically the frequency of cusps, one might not be able to assess very accurately whether two approaching curves actually cross each other or they are simply nearby; similarly pseudocusps can also arise if one tries to estimate the occurrence of cusps numerically because discretisation would generically generate grid approximations.  In addition, being interested in gravity waves emitted by the string's ongoing events such as cusps, it is important to also compute the gravitational signals emitted from any highly relativistic region of the string.

In order to investigate the occurence of cusps and pseudocusps on the string over a periodic non-dynamical evolution and the influence of several parameters on such occurence, we will study the average positions and standard deviation of ${\bf a'}$ and ${\bf -b'}$ on the unit sphere. We will then relate this probability to the string and network's parameters in order to determine the characteristics that can lead to cuspy events. Note that in the following, a ``cusp'' refers to either an actual cusps or a pseudo-cusp.

\subsection{Analytical considerations} \label{sec:analytic}

Here we define the $z$-axis as the axis of reflection that relates ${\bf a}'$ and ${\bf -b'}$, namely we align the $z$-axis with the $a'_\parallel$.
\begin{figure*}
	\begin{center}
	\includegraphics*[keepaspectratio, width=9cm, trim= 225 275 140 140, clip]{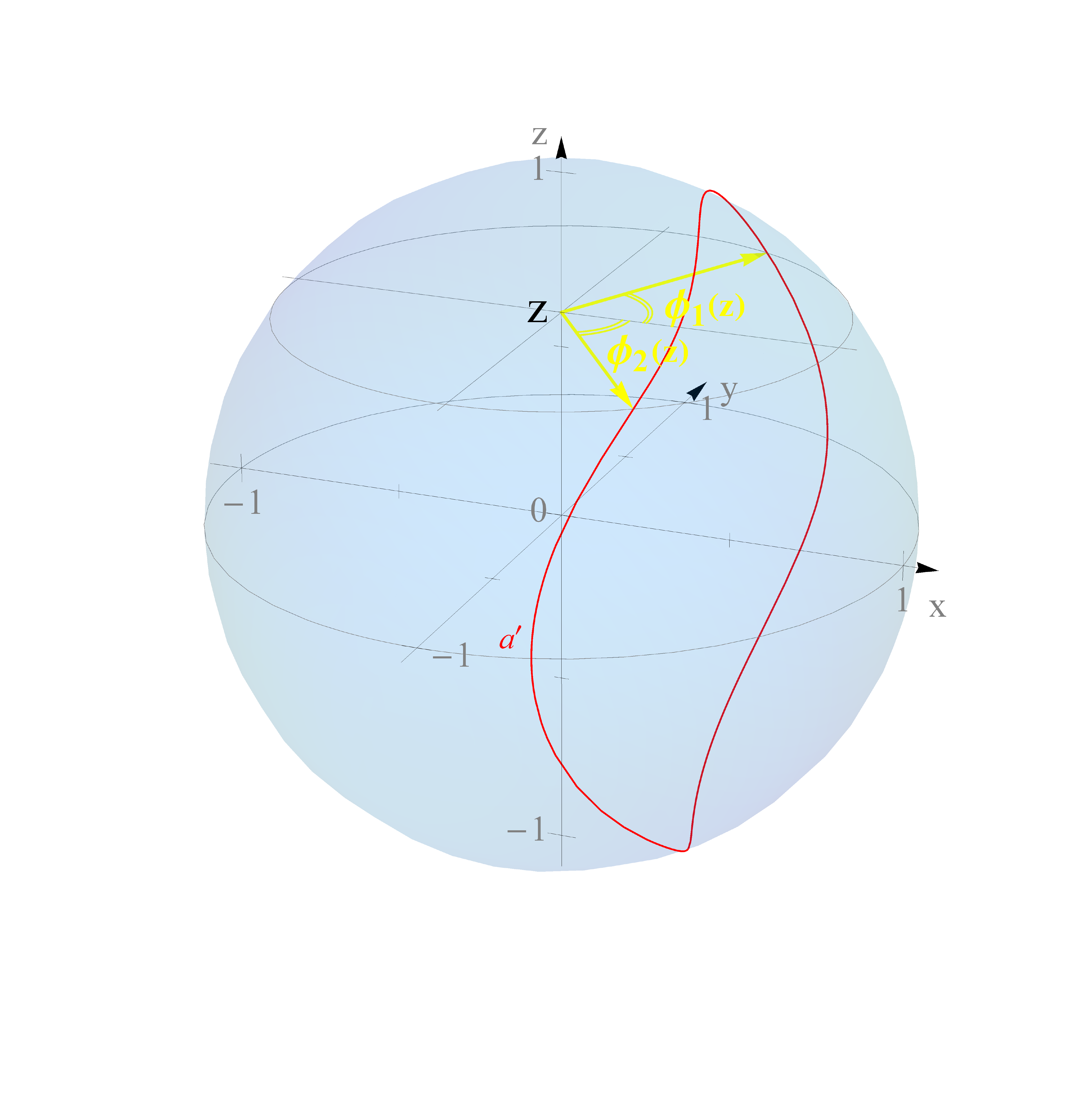}
	\caption{Cylindrical coordinates about the $z$-axis and the angles $\phi_i (z)$\\for the description of ${\bf a'}$ on the unit sphere.}
	\label{Reparametrisation}
	\end{center}
\vspace{-1em} \end{figure*}
Then the vectors ${\bf a}'$ can be written in cylindrical coordinates about this $z$-axis as in Fig.~\ref{Reparametrisation}, yielding
\be
	{\bf a}' = \left\{ \begin{array}{rcccl}
		\big( & \!\!\sqrt{1-z^2} \;\cos \phi_1 \left( z \right), & \!\sqrt{1-z^2} \;\sin \phi_1 \left( z \right) & \!\!,\,z & \!\!\big) \\
		\big( & \!\!\sqrt{1-z^2} \;\cos \phi_2 \left( z \right), & \!-\sqrt{1-z^2} \;\sin \phi_2 \left( z \right) & \!\!,\,z & \!\!\big)
	\end{array} \right.~,
\ee
for $z \in \left( z_{\rm min}, z_{\rm max}\right)$, where the two vectors come from the fact that most of the time the loop is at least double-valued in the $z$-coordinate. Cusps will appear whenever $\phi_1 \left( z \right) + \phi_2 \left( z \right) = \pi$, hence this is the condition we want to investigate. Let us define $2{\cal L}$ as the periodicity of the ${\bf a}'$ loop (which from the previous section needs not to be the same as the length $l$ of the string and can be different for different components). Hence,
\be \label{eq:ax}
	\langle a_x' \rangle_\sigma \equiv \frac{1}{2{\cal L}} \int^{\cal L}_{-{\cal L}} \!{\rm d} \sigma \;a_x' \left( \sigma+t \right)
	= \frac{1}{2\left( z_{\rm min} - z_{\rm max} \right)} \int^{z_{\rm max}}_{z_{\rm min}} \!\!{\rm d}z \;\sqrt{1-z^2} \left( \cos \phi_1 + \cos \phi_2 \right)~,
\ee
and
\be
	\langle a_y' \rangle_\sigma \equiv \frac{1}{2{\cal L}} \int^{\cal L}_{-{\cal L}} \!{\rm d} \sigma \;a_y' \left( \sigma+t \right)
	= \frac{1}{2\left( z_{\rm min} - z_{\rm max}\right)} \int^{z_{\rm max}}_{z_{\rm min}} \!\!{\rm d}z \;\sqrt{1-z^2} \left( \sin \phi_1 - \sin \phi_2 \right)~,
\ee
where we have dropped the explicit dependence of $\phi_i$ on $z$ for notational simplicity.

Similarly, we can write
\begin{subequations} \label{eq:variances}
	\begin{align}
	\langle a_x' a_x' \rangle_\sigma &\equiv \frac{1}{2{\cal L}} \int^{\cal L}_{-{\cal L}} \!{\rm d} \sigma \;a_x'\left( \sigma+t \right) a_x' \left( \sigma+t \right) \nonumber \\
	&= \frac{1}{2\left( z_{\rm min} - z_{\rm max} \right)} \int^{z_{\rm max}}_{z_{\rm min}} \!\!{\rm d}z \left( 1-z^2 \right) \left( \cos^2 \phi_1 + \cos^2 \phi_2 \right)~,\label{eq:varA}\\
	\langle a_y' a_y' \rangle_\sigma &\equiv \frac{1}{2{\cal L}} \int^{\cal L}_{-{\cal L}} \!{\rm d} \sigma \;a_y'\left( \sigma+t \right) a_y' \left( \sigma+t \right) \nonumber \\
	&= \frac{1}{2\left( z_{\rm min} - z_{\rm max} \right)} \int^{z_{\rm max}}_{z_{\rm min}} \!\!{\rm d}z \left( 1-z^2 \right) \left( \sin^2 \phi_1 + \sin^2 \phi_2 \right)~.\label{eq:varB}
	\end{align}
\end{subequations}
The sum of Eqs.~(\ref{eq:varA}) and (\ref{eq:varB}) leads to
\be \label{eq:variances2}
	\langle a_x' a_x' \rangle_\sigma + \langle a_y' a_y' \rangle_\sigma = \frac{1}{z_{\rm max} - z_{\rm min}} \int^{z_{\rm max}}_{z_{\rm min}} {\rm d}z \left( 1-z^2 \right)
	= \langle 1-z^2 \rangle_z ~,
\ee
thus providing a direct relationship between $\langle a_x' a_x' \rangle_\sigma$ and $\langle a_y' a_y' \rangle_\sigma$. Adding Eq.~(\ref{eq:variances2}) to the difference of Eqs.~(\ref{eq:varA}) and (\ref{eq:varB}), we get
\begin{align} \label{eq:axax}
	\langle a_x' a_x' \rangle_\sigma = \frac{1}{z_{\rm max} - z_{\rm min}} \int^{z_{\rm max}}_{z_{\rm min}} {\rm d}z
	\left(1-z^2\right) \Big[ &2 \cos^2 \left( \tfrac{\phi_1 + \phi_2}{2}\right) \cos^2 \left( \tfrac{\phi_1 - \phi_2}{2}\right) \nonumber \\
	- &\cos^2\left( \tfrac{\phi_1+\phi_2}{2}\right) - \cos^2\left( \tfrac{\phi_1-\phi_2}{2}\right) +1 \Big]~.
\end{align}
Let us consider the simplifying assumption $\phi_1\left( z \right) \approx \phi_2 \left( z \right)$, which we will later justify. Note that this condition means that the ${\bf a}'$ configuration is approximately symmetric through the $(xz)$-plane. Hence, Eq.~(\ref{eq:ax}) becomes
\be \label{eq:ave1}
	\langle a_x' \rangle_\sigma \approx \frac{1}{z_{\rm max} - z_{\rm min}} \int^{z_{\rm max}}_{z_{\rm min}} \!{\rm d}z \;\sqrt{1-z^2} \,\cos \left( \frac{\phi_1 +\phi_2}{2} \right)~,
\ee
whilst Eq.~(\ref{eq:axax}) reads
\be \label{eq:SD1}
\langle a_x' a_x' \rangle_\sigma \approx \frac{1}{z_{\rm max} - z_{\rm min}} \int^{z_{\rm max}}_{z_{\rm min}} \!{\rm d}z \left(1-z^2\right) \cos^2 \left( \frac{\phi_1+\phi_2}{2} \right)~.
\ee
Let us note that if the string is straight, the curve described by {\bf a'} is reduced to a point at the $x=1$ pole~; the further the string deviates from a straight line, the further the {\bf a'} curve will deviates from this pole. Only wavy strings could thus generate a curve that spans further than the $x>0$ half-sphere, that is further than the $(\phi_1, \,\phi_2) \in [0, \,\sfrac{\pi}{2}[^2$ half-sphere. Thus, the right hand side of Eq.~(\ref{eq:ave1}) is positive and it becomes smaller and smaller for wavier strings without changing sign. The condition we are interested in here is $(\phi_1 + \phi_2) \geq \pi$, since this would indicate that the curve described by {\bf a'} on the unit sphere spans over more than a whole half-sphere, implying a crossing with {\bf b'} by symmetry. Namely we would like to find the parameters for which there is a high probability that exists a $z \in \left( z_{\rm min}, z_{\rm max} \right)$ such that $\phi_1 \left(z\right) + \phi_2 \left(z\right) \geq \pi$, or equivalently such that
\be
	\cos \left( \frac{\phi_1 \left(z\right) + \phi_2 \left(z\right)}{2} \right) \leq 0~.
\ee
Noting that $|z_{\rm min}|\leq 1$ and $z_{\rm max} \leq 1$, we have $0 \leq 1-z^2 \leq 1$ for all $z \in \left( z_{\rm min}, z_{\rm max} \right)$ and hence we can rewrite the above condition as
\be \label{precond}
	\sqrt{1-z^2} \,\cos \left( \frac{\phi_1 \left(z\right) + \phi_2 \left(z\right)}{2} \right) \leq 0~.
\ee
The average of this quantity is given by Eq.~(\ref{eq:ave1}) and the fluctuations about this average are given by Eq.~(\ref{eq:SD1}). In particular, the standard deviation is
\be \label{eq:SD2}
	\sigma^2_{\left( \sqrt{1-z^2} \cos \left( \left( \phi_1 + \phi_2 \right) /2 \right) \right) } \approx \langle a_x' a_x' \rangle_\sigma - \langle a_x' \rangle_\sigma^2~.
\ee
Thus, we have the average (which is positive) and the standard deviation of a quantity, for which we want to calculate the probability to be somewhere negative. This is likely to happen if the standard deviation is larger than a significant fraction of the average. This means that the probability of the quantity of interest being negative is significant when
\be
	\alpha \;\sigma^2_{\left( \sqrt{1-z^2} \cos \left( \left( \phi_1 + \phi_2 \right) /2 \right) \right)}
	\gtrsim \left \langle \sqrt{1-z^2} \cos \left( \,\sfrac{(\phi_1+\phi_2)}{2} \,\right) \right \rangle^2_x~,
\ee
with $\alpha$ being between $1$ and $5$. It corresponds to a few times the standard deviation being larger than (or comparable to) the average. To illustrate the idea, let our quantity $a_x'$ follow a gaussian distribution; then, for instance $\alpha=2$ would mean that a string should present a significant number of cusps if Eq.~(\ref{precond}) was satisfied for about $2.5\%$ of the points on the string --- $2\sigma$ corresponding to a $95\%$ confidence level.

Thus, using Eqs.~(\ref{eq:ave1}) and (\ref{eq:SD2}) we find that there is a significant probability of having cusps provided
\be \label{eq:criterion}
	\langle a_x' a_x' \rangle_\sigma \gtrsim \frac{1+\alpha}{\alpha} \left( \frac{|{\bf \Delta}|}{\sigma_{m}} \right)^2 = \frac{1+\alpha}{\alpha} \;\Delta_a^2~,
\ee
where we have used that ${\bf a}'$ is periodic in $2{\cal L} = 2\sigma_{m}$ (from Eq.~(\ref{eq:periodz})), defined
\begin{align}
	{\bf \Delta} &= \left( \Delta, \,0, \,0 \right) \equiv {\bf x} (\sigma_{m},t) - {\bf x} (0,t)~, \\
	\Delta_a &\equiv \frac{1}{2\sigma_{m}} \int^{\sigma_{m}}_{-\sigma_{m}} {\rm d} \sigma \,a_x' \left( \sigma+t \right)~,
	\quad \Delta_b \equiv \frac{1}{2\sigma_{m}} \int^{\sigma_{m}}_{-\sigma_{m}} {\rm d} \sigma \,b_x' \left( \sigma+t \right)~,
\end{align}
and used the relations $\Delta_a = - \Delta_b$ and $\Delta /\sigma_{m} = (\Delta_a - \Delta_b)/2 = \Delta_a$. This is a key result as it gives a simple way to discriminate between cuspy and non-cuspy strings, simple in the quantities to compute and in the physical meaning behind inequality (\ref{eq:criterion}).

The prefactor $\sfrac{(1+\alpha)}{\alpha}$ lies somewhere between $1$ and $2$, the latter being too conservative (it corresponds to $\alpha = 1$, meaning there should be cusps only if more than $15 \%$ of the curve satisfy Eq.~(\ref{precond})) and the former not constraining enough (where $\alpha \gg 1$, that is a very small fraction of the curve satisfying Eq.~(\ref{precond}) is sufficient to generate cusps along the string).

Note that the approximation $\phi_1 \left( z \right) \approx \phi_2 \left( z \right)$ can be easily satisfied when looking at the string with a probabilistic point of view. Indeed, one can continuously deform the curve ${\bf a'}$ to get a symmetric curve with respect to the $(xz)$-plane. If this transformation conserves the statistical description of the curve, it does not change significantly the probability of the curve to intersect its image under the symmetry with respect to the $z$-axis. What should be conserved in the transformation is only the proportion of the curve reaching a certain distance to its mean position. It is possible to continuously deform our curve maintaining such properties, especially if we are looking at a large population of strings in which tiny variations on each string are smoothed over the number of them.

Recall that we have defined the $z$-axis so that the heavy string at the $\sigma =0$ junction is aligned along this $z$-axis. Equation~(\ref{eq:criterion}) implies a minimum distance reached by the $x$-component of ${\bf a'}$ (and ${\bf - b'}$) from its average circle, defined as the circle in the $(yz)$-plane whose center $C$ is at a distance $\Delta_a$ from the centre of the sphere on the $x$-axis. This equation can be also understood as implying a boundary on how irregular the velocity of the two movers have to be to generate a substantial amount of cusps.

In order to make a link with the string network's and the individual string's parameters, let us first recall that $\Delta$ is the distance between the two ends of the string, stretched between the two junctions. Rescaling $\Delta$ by the parameter length of the string $\sigma_{m}$, this gives the distance in the unit sphere between the two average circles for ${\bf a'}$ and ${\bf -b'}$. At a fixed length, if $\Delta$ increases, the two circles are shifted away and the probability of cusps decreases; at fixed $\Delta$, if the length increases, the cumulated length of the curve's parts reaching the minimum distance increases too so the number of cusps becomes larger. Hence, the number of cusps is lower for straighter strings.  Moreover, if the string has large-amplitude waves, the curves ${\bf a'}$ and ${\bf -b'}$ deviate from their average position and the number of cusps increases. Hence, strings with large waves are expected to have more cusp events. At a fixed length, if the curves have less large waves, they will exhibit a larger amplitude and thus there will be more cusps. So, a long string with large-amplitude waves should exhibit more cusps than a short straight string or a small-scale structured string.

Recall this is a qualitative analysis of a non-dynamical non-interacting string with \mbox{Y-junctions} and let us emphasise that the aim here is to estimate the number of cusp events. Still, it is important to identify the relevant parameters in such setups and to understand their influence. This will be done in more details in the following analysis presented mainly in Section~\ref{sec:numsim} and linked to some of the usual network and string parameters in Section~\ref{sec:correlparam}.

\subsection{Pseudocusps and velocity}

Let us recall that a pseudo-cusp is defined as a point at which the left- and right-movers' \emph{3}-velocity vectors $\bf{a'}$ and $\bf{b'}$ are very close to each other, enough for the point to be highly relativistic, but not exactly equal to each other. We define $\sigma^{\mathrm{clos.}}_\pm = \sigma^{\mathrm{clos.}} \pm t^{\mathrm{clos.}}$ to be the null coordinates for which these two vectors are the closest in this neighbourhood, and denote by $\theta_c$ the angle between the two vectors at $\sigma^{\mathrm{clos.}}_\pm$. We also denote
\begin{align}
	l^\mu = \dot x^\mu (\sigma^{\mathrm{clos.}}, t^{\mathrm{clos.}}) = \sfrac{1}{2} \, (a^\mu (\sigma^{\mathrm{clos.}}_+) - b^\mu (\sigma^{\mathrm{clos.}}_-)) \\
	\mathrm{and} \qquad \delta^\mu = \sfrac{1}{2} \, (a^\mu (\sigma^{\mathrm{clos.}}_+) + b^\mu (\sigma^{\mathrm{clos.}}_-))
\end{align}
the half-sum and the half-difference between the left- and right-movers' \emph{4}-velocities, respectively. Note that, despite what it looks like, we here call $l^\mu$ the half-\emph{sum} recalling the vectors we are interested in are $a'^\mu$ and $- b'^\mu$.

The \emph{4}-vector $l^\mu$ is the \emph{4}-velocity at the point of interest and we recall that it is a null vector in the cusp case. In the case of pseudocusps, the time-component $l^0$ is also equal to $1$, but the norm of the \emph{3}-velocity of the string at that point $(\sigma^{\mathrm{clos.}}_+, \sigma^{\mathrm{clos.}}_-)$ equals
\be \label{eq:1-theta2}
	|l^i| = \sqrt{\frac{1 + \cos(\theta_c)}{2}} \approx 1 - \sfrac{\theta_c^2}{8}~;
\ee
however, $\delta^\mu$ is space-like, with $\delta^0 = 0$ in the time gauge, and
\be
	|\delta^i| = \sqrt{\frac{1 - \cos(\theta_c)}{2}} \approx \sfrac{\theta_c}{2}~.
\ee
The angle $\theta_c$ can be thought of as measuring the \emph{softness} of a relativistic part of the string. The larger it is, the smaller the velocity and the softer the pseudo-cusp; for $\theta_c = 0$, the event is an actual cusp and the velocity reaches $c=1$.

We would also like to evaluate the number of pseudocusps statistically. The problem has to be looked at using the unit sphere description along with the mean and standard deviation of the curves drawn by ${\bf a'}$ (and ${\bf - b'}$). Let's first recall that a pseudo-cusp is related to the ${\bf a'}$ curve approaching its symmetric counterpart without crossing it, while a cusp is linked to the curve crossing its counterpart. Let's then define the relative distance between the curves as a positive number when the curves remain in their natural half-sphere, becoming negative when the curves cross each other (i.e. between two crossings). One can then relates every pair of cusps and every pseudo-cusp to a minimal value of the distance: if this minimum is positive the string presents here a pseudo-cusp, and a pair of cusps if it is negative.

In addition, below the mean value of this relative distance, the lower the distance, the smaller the proportion of the curve reaching such a distance. Still, a minimal distance being small and positive happens roughly as often as a minimum being small and negative. This implies that a pseudo-cusp should appear as often as a pair of {\it narrow} cusps; we here define narrow cusps as a pair of cusps for which the minimal distance reached is small and negative (in opposition to what could be called {\it large} cusps, for which the minimum distance becomes large and negative between the two cusps).

In terms of the relative occurence of cusps and pseudocusps, one can deduce that a string with cusps should also present pseudocusps. In addition, since {\it large} cusps are rarer than {\it narrow} cusps, there should be a bit more than twice as many cusps as there are pseudocusps, approximately.

\section{Numerical simulation} \label{sec:numsim}

\subsection{Method}

We develop a simulation of the previously described configuration in order to check the considerations made and to evaluate the occurence of cusps and pseudocusps. Our code depends on both the string network's and the individual string's parameters --- namely $\xi$ and $\bar \xi$, as we will see below --- and is based on the following assumptions. Firstly, the string's ends are fixed on the heavy strings, being themselves insensitive to the motion of the light string and to any transfer of momentum.  In addition, the quasi-periodic cases are neglected and the position and velocity of the string at $t = 0$ are defined by a Fourier series (i.e. by the amplitude of each mode). These amplitudes are all drawn in $[-h_m,\,h_m]$,\footnote{A uniform distribution in the interval $[-h_m,\,h_m]$ has been initially encoded. Note though that there is a bias: indeed, high values of the amplitude imply high velocities, i.e. more strings whose parts may travel faster than $c=1$ --- which is obviously forbidden. These strings are dismissed immediately, distorting a posteriori the uniform draw within the interval.} where $h_m$ is a prefixed highest value and the modes are the $n$ first harmonics of the string (up to $n$ nodes); with $n$ and $h_m$ being parameters of the simulation. More precisely, they set up the oscillatory behaviour of the string, fixing a limit to the highest frequency and to the amplitude reached in its Fourier decomposition.

The parameter length $\sigma_m$ and the interstring distance $\Delta$ are also inputs in the simulation. Indeed, to geometrically set up the system, one needs the end-to-end distance; additionally, the parameter length of the string is related to the fundamental frequency and to how wavy or wiggly the string can be. Clearly, $\Delta$ bounds $\sigma_{m}$, since the string cannot be shorter than the distance between its end points; one can also see that for $\sigma_{m} \rightarrow \Delta$ (and $\sigma_m > \Delta$), the curves ${\bf a'}$ and ${\bf -b'}$ get confined away from each other in the pole regions and ultimately shrink to a point in the case $\sigma_m = \Delta$. Since we will be mainly interested in their ratio, we chose to fix $\Delta$ by assigning to the end points invariable coordinate triplets while promoting $\sigma_{m}$ as one of the main parameters of the code.

\begin{figure}[t]
	\begin{center}
	\includegraphics*[keepaspectratio, width=9cm, trim= 142 300 70 275, clip]{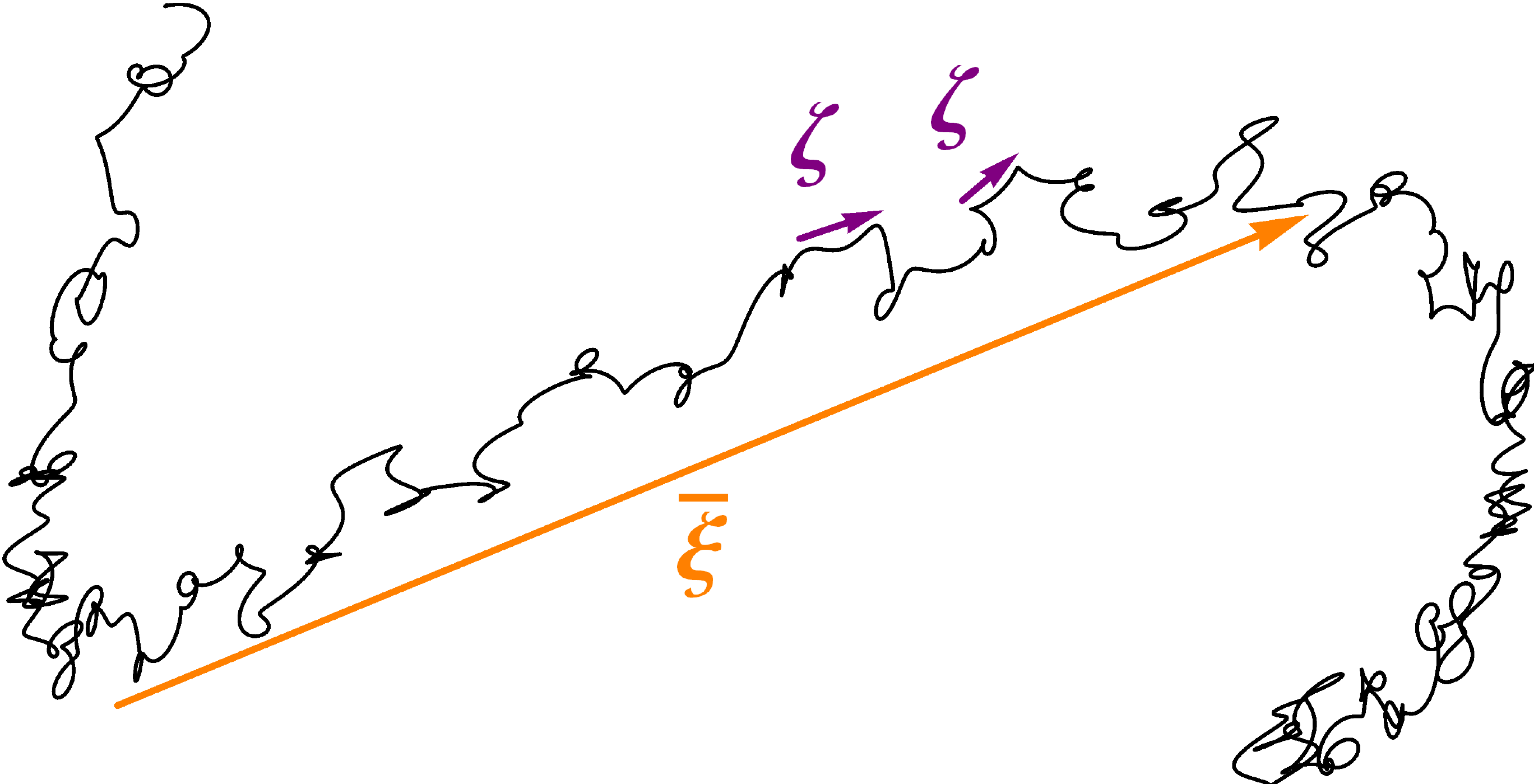}
	\caption{$\bar \xi$ and $\zeta$, two of the network's length scales.}
	\label{Lengthscales}
	\end{center}
\vspace{-1em} \end{figure}

The network's parameters are often chosen to be $\xi$, $\bar{\xi}$ and $\zeta$, representing the average interstring distance in the network, the coherence length scale (or large-scale structure) and the wiggliness (or small-scale structure)~; see for instance, Ref.~\cite{Austin:1993rg}). Equivalently, $\zeta$ is related to small wiggles and to edgy bends on the string, while $\bar \xi$ characterises large-amplitude waves. We denote by {\it ripple} both of these variations along the string, wiggles and wiggliness being related to the small-scale structure and thus to $\zeta$, while (large-amplitude) waves and waviness refer to the large-scale structure, that is, to $\bar \xi$. Fig.~\ref{Lengthscales} gives a schematic representation of these $\zeta$ and ${\bar \xi}$ length scales.

In our simulation, $\Delta$ can be identified as the distance $\xi$\footnote{We here consider for simplicity an overall interstring distance $\xi$ --- and generally only one set of parameters. As discussed in Section~\ref{sec:correlparam}, one can also consider that the light string and the heavy string networks have different characteristics, leading to the definition of $\xi_{light}$ and $\xi_{heavy}$. In such a scenario, $\Delta$ would be related to $\xi_{heavy}$ only.} between two heavy strings, even though what matters here is the ratio $\Delta / \sigma_m$. Note that this ratio could also be related to the large- and small-scale structure since a longer string has to exhibit more ripples, whatever the size of these ripples is. Here, there is no small-scale structure strictly speaking since the number of modes is quite low. So the wiggliness $\zeta$ is not defined and its influence is therefore not addressed.  In addition, there is no clear input for the large-scale structure and its characteristic length $\bar \xi$ is to be linked with several other parameters such as the number and amplitude of the vibration modes at $t=0$ or during a period. A crude estimation could be a fourth of a wavelength of the highest frequency mode present on the string, that is $\bar \xi \sim \,\sfrac{\sigma_m}{\,2 \bar n}$, where $\bar n$ is the highest frequency mode on the string (and not the input $n$, which is only a bound on the highest possible mode). One could also consider the amplitude of the waves, for instance estimating the standard deviation of the $y$- and \mbox{$z$-components} of the position of the string at $t=0$. The geometric mean of these two figures would represent even more accurately the characteristic size of a wave on the string, taking into account the two directions of extension of such large-amplitude waves.

Among the other ways to evaluate how wavy the string is, is to use the standard deviation of the $x$-component of the left- and right-movers' velocities, namely $\langle a'_x a'_x \rangle - \langle a'^2_x \rangle$ (and the same with $b'_x$) since it quantifies how far and how often the string goes away from a straight(er) position. Indeed, the straight line is represented by a constant ${\bf a'}$ and ${\bf -b'}$, while a large standard deviation from this pointlike curve means strong variations in the movers' amplitudes and smaller radii of curvature along the string.

Our simulation thus starts from these assumptions and parameters and a significant number of different string configurations is simulated. Each string's (non-interacting) evolution is then computed over a period. The string is then decomposed in a large number of points (each of them corresponding to a segment in our numerical simulation) and the period is decomposed in time lapses. We thus obtain a velocity distribution and its evolution over the period.  The number of cusps is found by analysing the curves on the unit sphere and looking for actual crossings; the velocity is then computed and checked to reach $c = 1$ within the numerical uncertainties --- which are generally\footnote{We found about $10\%$ of the cusps with velocities outside a $10^{-6}$-wide band around $1$, and $3\%$ outside a $10^{-5}$-wide band.} below $10^{-6}$. The pseudocusps are all the other highly relativistic areas; here, we consider as ``highly relativistic'' any velocity above $0.999 \,c$. Note that pseudocusps velocities are in a vast majority\footnote{We found more than $80\%$ of the pseudocusps' velocities below $1-10^{-5}$ and about $90\%$ below $1-10^{-6}$. Figures are presented here for the computed velocity.} in the range $[1-10^{-3},\,1-10^{-6}]$, helping to split between cusps ($1-v < 10^{-6}$) and pseudocusps ($10^{-6} \le 1-v \le 10^{-3}$).  Finally, it is checked that pseudocusps correspond to configurations with a very small gap between the two curves on the unit sphere; the angle $\theta_c$ between the two vectors ${\bf a'}$ and ${\bf -b'}$ is computed and its minimum found (within the grid approximation).

Even though our analysis is performed within a specific setup, our qualitative results remain valid in the more realistic string configurations. The slow motion of the heavy strings can be ignored as compared to that of the light strings, whilst the periodicity can be safely considered as generic. The absence of a dynamical analysis and interaction between strings, chosen for the simplicity of the computations, should not modify the way the network parameters influence the occurence of cusps and pseudocusps. In conclusion, our setup could represent a network  of heavy and light strings interacting at a time scale which is not to small compared to the period of the light string's movement. Hence, the correlation between the network parameters and the occurence of cuspy events should be valid independently of whether our simplifying assumptions are relaxed or not. Appendix~\ref{sec:snapshots} presents some example snapshots of a simulated string.


\subsection{Description of pseudocusps}

In the following, we call \emph{computed} velocity the one from the simulation's direct evaluations, namely the highest velocity locally reached as it has been computed, and \emph{theoretical} velocity the value obtained using our model of pseudocusps,
\begin{figure*}[t]
	\begin{center}
	\includegraphics[height=7.5cm, width=13.05cm]{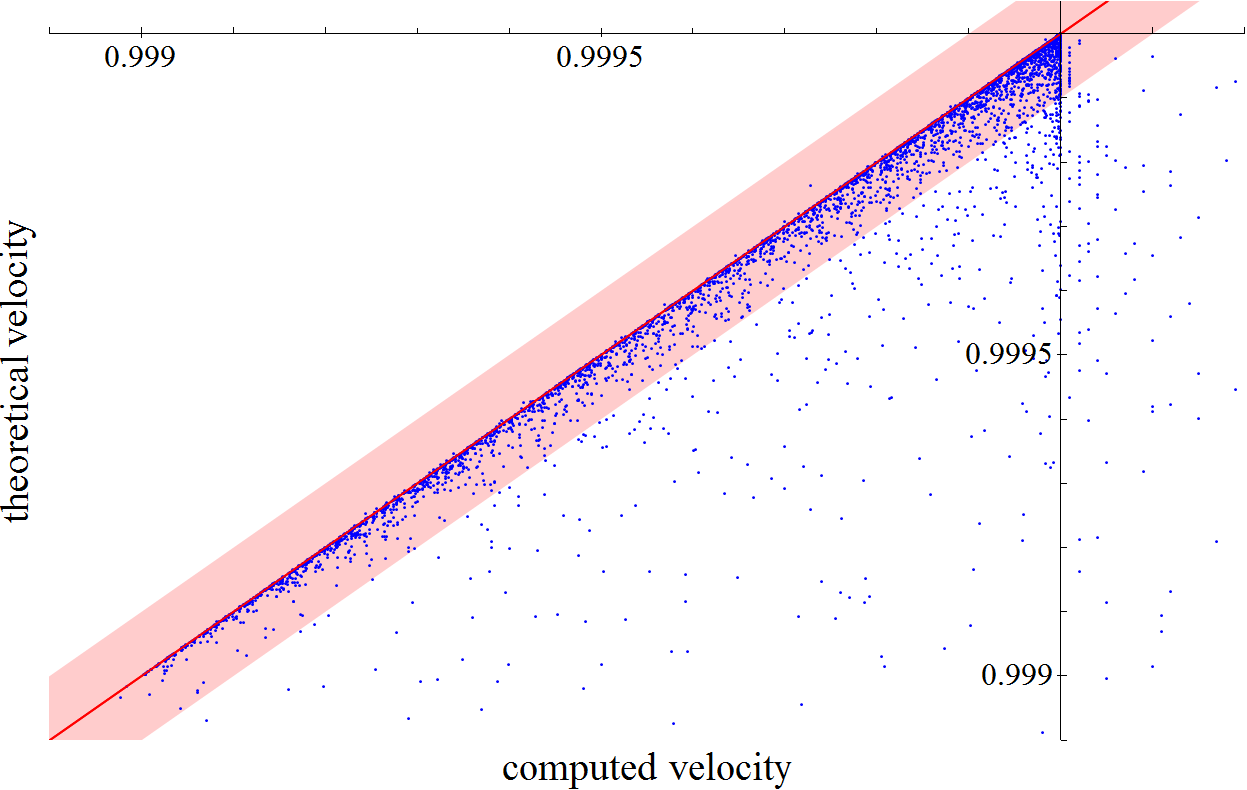}
	\caption{Pseudocusps: theoretically estimated velocity versus computed velocity.\\
		Note that $80\%$ of the pseudocusps present a difference between the two velocities below\\
		$10^{-4}$, meaning that it is represented here by a point in the red shaded area.}
	\label{PCVelocityAngle}
	\end{center}
\vspace{-1em} \end{figure*}
namely the one we got using the approximation $(1\,-\,\sfrac{\theta_c^2}{8}\,+\,\sfrac{\theta_c^4}{384})$ from Eq.~(\ref{eq:1-theta2}).\footnote{The approximation used here takes into account one more term, even if it is very often insignificant compared to the numerical uncertainties.} One can note that the latter cannot be above $1$. We obtain that there is a very good agreement between these two estimations of string the velocity at the pseudocusps.

Fig.~\ref{PCVelocityAngle} shows, for almost 4300 pseudocusps\footnote{About $8\%$ of the almost 4700 pseudocusps studied here are not represented on this plot.} the computed velocity versus the theoretically estimated one. The red line draws the equality case and one can immediately note that $v_{\rm th} \le v_{\rm cp}$ (except in a very few cases almost not visible on this plot). This is probably due to the methods used: in the first case, the velocity has to be above $0.999$ whereas in the second one it is always below $1$. In addition, the computed velocity is subject to quite a lot of grid and computational uncertainties and can thus reach $1$ (or even a higher value) fairly easily.\footnote{We found almost   $10\%$ of the pseudocusps' computed velocity above $1 + 10^{-6}$. Recall that our uncertainties are generally of the order of $10^{-6}$.} Finally, more than $80\%$ present a difference between the two velocities which is below~$10^{-4}$.

Note though that all these discrepencies are actually gathering on the same cases. Indeed, among the $6\%$ pseudocusps with theoretical velocity below $0.999\,c$, $80\%$ give a computed velocity above $1-10^{-6}$. Also, almost $60\%$ of the pseudocusps presenting velocities' discrepancies larger than $10^{-4}$ have either an abnormally small theoretical velocity or an abnormally large computed velocity.

\subsection{Occurence of cusps and pseudocusps}

In order to check if the criterion set up in Eq.~(\ref{eq:criterion}) is actually discriminating between configurations with cuspy phenomena and those without any cusp or pseudo-cusp, we simulated and studied a significant number of strings ($237$) within a variety of parameters. From the curves ${\bf a'}$ and ${\bf -b'}$ have been calculated both the number of cusps and pseudocusps and the mean and standard deviation of ${\bf a'}$ in the $x$-direction. A very good agreement has been found between the presence of cuspy phenomena and the completion of our criterion.

\begin{figure*}[t]
	\begin{adjustwidth}{-2em}{-2em}
	\includegraphics[height=9.84cm, width=16cm]{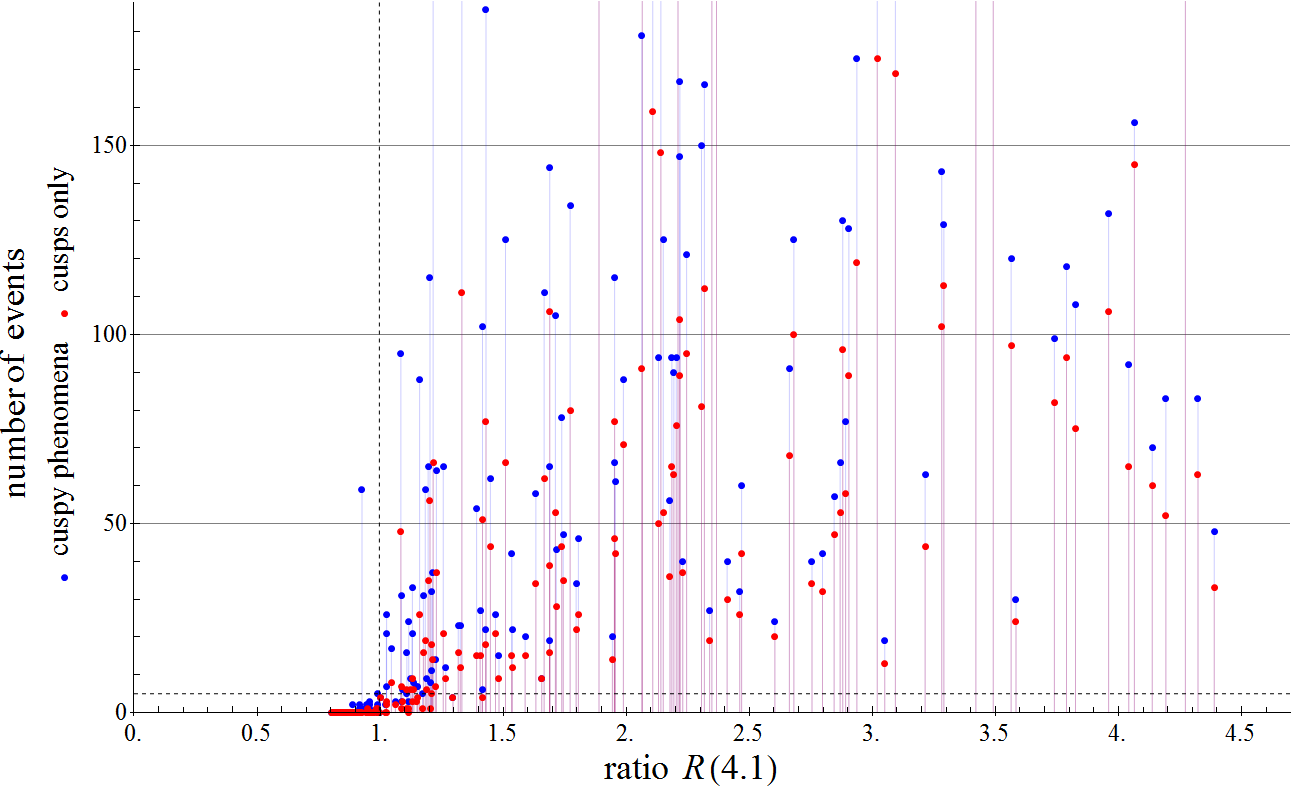}
	\caption{Number of cusps (red) and cuspy phenomena (blue) vs. ratio $\sfrac{\langle a_x' a_x' \rangle}{\,1.24 \:\Delta_a^2 \,}$.\\
		The black dashed line standing at $R=1$ is splitting the plane in two parts:\\
		non-cuspy strings for low ratios and cuspy strings for high ratios.}
	\label{nbCPvsRatio}
	\end{adjustwidth}
\vspace{-0.8em} \end{figure*}

On Fig.~\ref{nbCPvsRatio} we plot the number of cuspy phenomena versus the ratio
\be \label{cuspRatio}
	R \,(\alpha=4.1) \equiv \left. \frac{\langle a_x' a_x' \rangle}{\frac{\alpha+1}{\alpha}\:\Delta_a^2 \,} \,\right|_{\alpha=4.1}
	= \frac{\langle a_x' a_x' \rangle}{\,1.24 \:\Delta_a^2 \,}~,
\ee
where the constrain parameter $\alpha$ can take any arbitrary value. Here it has been {\sl a posteriori} fixed to $4.1$, for convenient reasons we will explain below. Recall that once $\alpha$ is fitted, we are expecting to have only strings with no cusps or pseudocusps for a ratio $R(\alpha) < 1$, and strings with cuspy phenomena for $R(\alpha) > 1$. Phrased differenrently, we should have neither non-cuspy strings with $R(\alpha) > 1$, nor cuspy ones with $R(\alpha) < 1$.

Note though that our statistical approach --- both from the definition of the ratio $R(\alpha)$ and from the number of strings considered --- will probably lead to strings in the tail of the distribution. Indeed, even with the most reliable choice of $\alpha$, we are expecting to find a small range of value around $1$ for which there are both strings with and without cuspy phenomena. If such an interval around $1$ is not too large, this is not in contradiction with our previous analysis and does not affect the coherence of the results presented here.

Each simulated string is represented by two aligned\footnote{Since the two dots stand for the same string, the ratio on the horizontal axis is the same.} dots: we use the red one to read on the vertical axis the number of cusps, and the blue one for the number of cuspy phenomena (both cusps and pseudocusps). The shaded coloured vertical lines are guides to read and have no physical meaning; it also helps tracking points whose vertical coordinate is off the plotted range.
The choice of the value of $\alpha$ and of where we divide the plane in two has to be discussed in view of the results. Before getting into the details, one can notice that the chosen value indeed fits with our set of points: on the left of the black dashed line standing at $R=1$ are mainly non-cuspy strings, while on the right one we can almost only find cuspy strings. In addition, as we foreseen the range in which one can find both behaviours is restricted --- roughly between $0.9$ and $1.1$. This means that strings satisfying the inequality
\be \label{finalRatio}
	R \,(\alpha=4.1) \gtrsim 1 \quad \Leftrightarrow \quad \langle a_x' a_x' \rangle \gtrsim 1.24 \,\Delta_a^2 \,
\ee
would generally present cusps, and vice-versa.

To be more accurate, let us zoom on what is happening around $0.9$--$1.1$ and let us discuss the ways to draw the limiting ratio. One may note that different rules can be set up to cut the plane in two parts (one without and another one with cusps).  Firstly, one can decide to look at the highest ratio associated with a string presenting no cuspy events in order to fix the separating ratio (let's call it the \emph{Highest with No Cuspy Events} ratio, i.e. the HNCE). One can also consider the string with the lowest ratio and at least one cusp or pseudo-cusp (giving the \emph{Lowest With Cuspy Events} ratio, or LWCE). Note that since the HNCE is higher than the LWCE, there is a ratio interval in which we found both strings with and without cusps --- again, as was expected.  Alternatively, one can choose to look at cusps only and follow the same method, giving two other boundary ratios (namely the HNC and the LWC, ``C'' standing for \emph{Cusp(s)}). Note that these two new values are higher than their cuspy phenomena counterparts as pseudocusps are more likely to happen than cusps for borderline configurations. One thus gets four different ratio values which can equally be considered as valid turning points. One also has two intervals within which cuspy phenomena and cusps appear.
\begin{figure*}[t]
	\begin{adjustwidth}{-6em}{-5em}
	\begin{center}
	\begin{subfigure}{9cm}
	\includegraphics[height=5.5125cm, width=8.75cm]{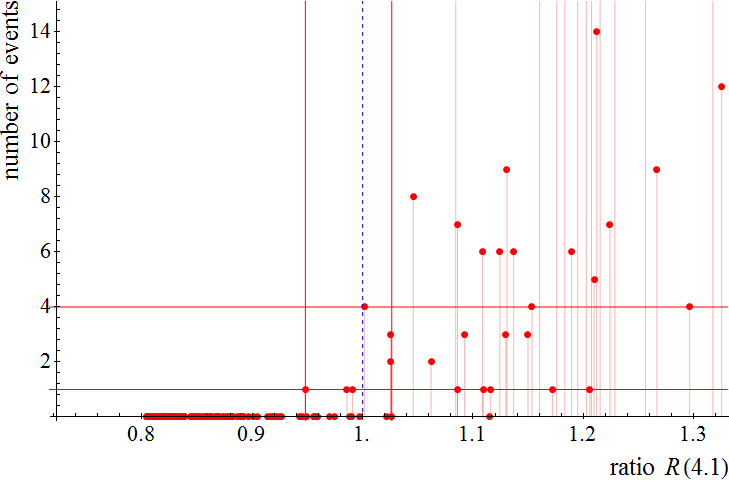}
        \setlength{\abovecaptionskip}{-1pt}
	\caption{cusps only}
	\label{nbCPvsRatioCU}
	\end{subfigure}
	\begin{subfigure}{9cm}
	\includegraphics[height=5.5125cm, width=8.75cm]{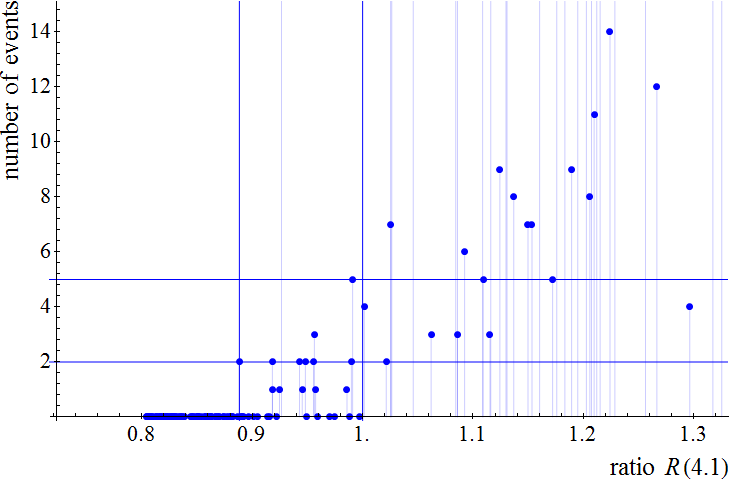}
        \setlength{\abovecaptionskip}{-1pt}
	\caption{cusps and pseudocusps}
	\label{nbCPvsRatioCP}
	\end{subfigure}
	\caption{Zoom around the low numbers of cuspy events.\\
		The vertical lines mark where the different splitting rules divide the plane.}
	\label{nbCPvsRatioZoom}
	\end{center}
	\end{adjustwidth}
\vspace{-1em} \end{figure*}

Depending on which rule one decides to apply, one gets a different line splitting the plane, giving a different value for $\alpha$. Again, this is nothing to worry about since we obtained quite close values, between $0.9$ to $1.1$.\footnote{We decided to neglect the two strings (over $237$) presenting exceptional behaviours: one with no cusp and a quite high ratio --- compared to the second-highest ratio for a string with no cusp --- and one with a very large number of pseudocusps but a low ratio and no cusp. They are thought to be statistically irrelevant.} In each of the two in-between intervals, we obtained strings with a small number of cuspy phenomena: less than $4$ cusps or less than $5$ pseudocusps. Also, for larger ratios, we only get a very few strings presenting so few cuspy phenomena and these have all reasonably small ratios. These results confirm the expected behaviour apart from the exceptional strings lying in the tail of the distribution and thus not giving the typical response are within an anticipated range.

Fig.~\ref{nbCPvsRatioZoom} focuses on the bottom left corner of Fig.~\ref{nbCPvsRatio}\footnote{Again, the shaded coloured lines connecting points are guides for reading and help tracking points off the plot.} and has been divided in two plots: on the left and in red, Fig.~\ref{nbCPvsRatioCU} shows the number of cusps only versus the ratio $R(4.1)$ and on the right and in blue, Fig.~\ref{nbCPvsRatioCP} does the same for all cuspy events. On each of them, two of the four aforementioned ratios are represented by solid coloured lines: two red lines for the LWC and the HNC on Fig.~\ref{nbCPvsRatioCU} and two blue ones for the LWCE and the HNCE on Fig.~\ref{nbCPvsRatioCP}. Note that on Fig.~\ref{nbCPvsRatioCU} is also displayed a blue dashed line marking the HNCE ratio (i.e., the highest of the two ratios for all cuspy phenomena); it is lying roughly in the middle of the interval considering cusps only (on the graph, the two solid red lines).

We would like to determine a value for the ratio which splits the plane in two regions (without and with cuspy phenomena), knowing that in a small neighbourhood around this value one should expect to find irregularities, which we expect to be sufficiently rare and small. One can see that the HNCE ratio satisfies our needs:
\begin{itemize} \itemsep-1.1em \vspace{-0.55em}
	\item on the left (i.e., for smaller ratios than the value of the HNCEevents --- most of them presenting no cusp and no pseudo-cusp;\newline \parskip0em
	\item on the right (i.e., for higher ratios) lies only strings with at least two cusps and pseudocusps, most of them presenting more than three cusps and five cuspy phenomena.
\end{itemize} \vspace{-0.6em}

In addition, recall that our analytic work to find the ratio $R(\alpha)$ is identifying cusps and pseudocusps (see Section~\ref{sec:analytic}), so the most meaningful turning point values we found are the ones related to all cuspy phenomena (HNCE and LWCE). Hence, the choice we made at the beginning to set $\alpha=4.1$.

We have set up here a quick and efficient method to discriminate between cuspy strings and non-cuspy ones.

\subsection{Number of cusps and pseudocusps}

One can now try to find which parameters influence the number of cusps and pseudocusps on a string. As we have seen already, there is a strong dependence on the interstring distance $\Delta = \xi$ and the parameter length of the string $\sigma_m$ --- or rather on $\sfrac{\Delta}{\sigma_m}$ --- as well as some important correlation with the mean squared \emph{x}-component of the string's movers' velocities $\langle a_x' a_x' \rangle$ and $\langle b_x' b_x' \rangle$.
\begin{figure*}[t]
	\begin{center}
	\hspace{-2em} \includegraphics[keepaspectratio, width=15cm]{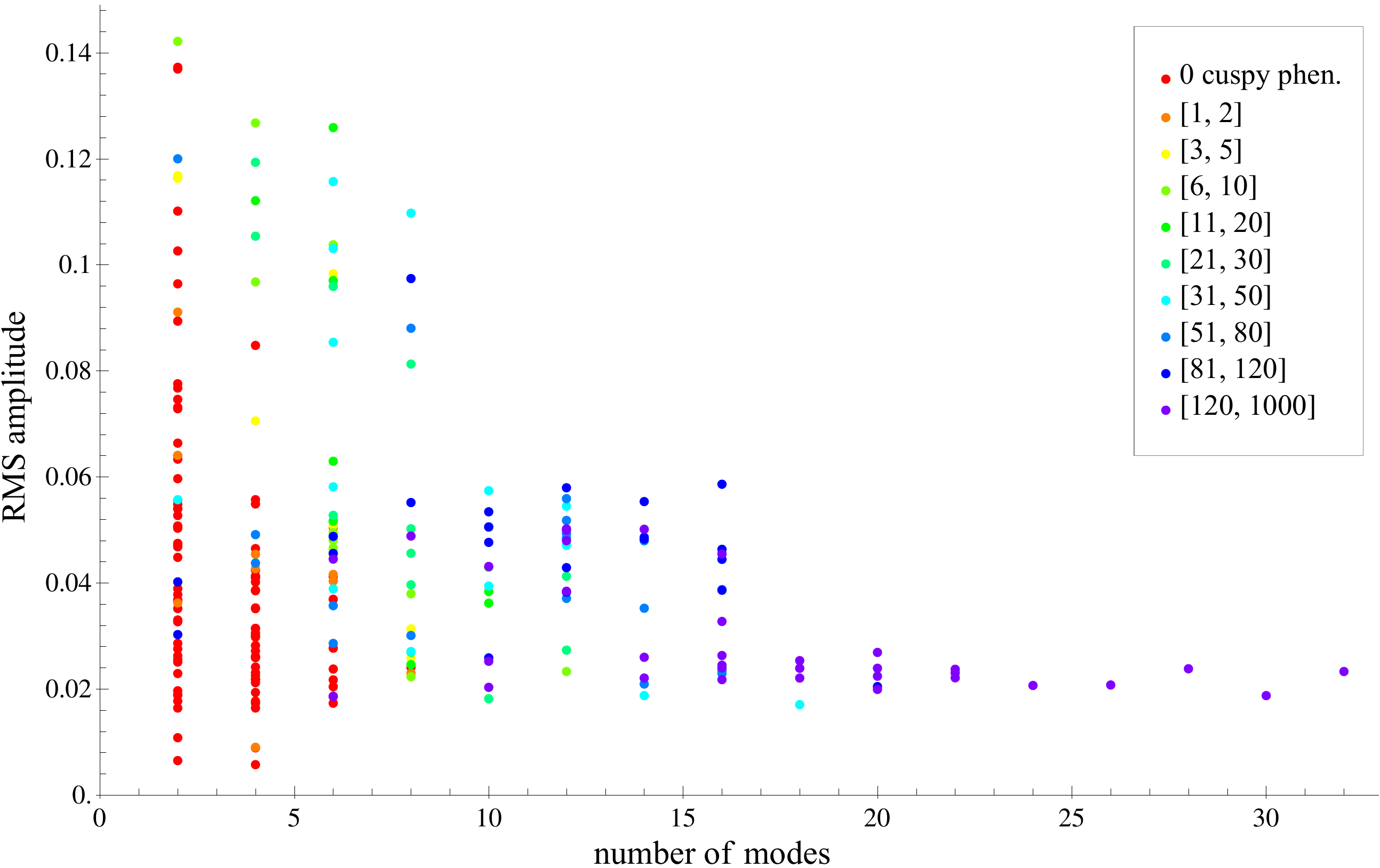}
	\caption{Root mean square amplitude of the $x$-modes versus the number of $x$-modes.\newline
		From red to purple, strings with $0$ to between $120$ and $1000$ cuspy events.}
	\label{XmodesColCP}
	\end{center}
\vspace{-1.5em} \end{figure*}

In order to understand these relations in more detail, we first analyse the influence of the Fourier modes initially implemented in the string and found that only the $x$-modes\footnote{The $y$- and $z$-modes are not found to be correlated to the number of cusps. The number and amplitudes of these modes are only indirectly linked to those of the x-modes via the fact that $(a'_{\mu})^2 = 1$.} influence the number of cusps, both via the number of modes and their amplitudes. On Fig.~\ref{XmodesColCP}, we plot the root mean square of the amplitudes versus the number of modes; a colour gradient is representing the strings grouped according to the number of cuspy events (from $0$ in red to above $120$ in purple). It is first obvious that more modes imply a lower RMS amplitude. This is due to the physical constraint to have no supraluminal points on the string.\footnote{This constraint is enforced during the evolution of the string but has to be carefully checked at $t=0$.} In addition, one can note that a low number of $x$-modes implies a low number of cusps, especially for low RMS amplitudes. Also, many modes generate strings with statistically many more cusps. For a fixed number of modes, higher amplitudes are associated with strings with more cusps, whereas at a fixed RMS amplitude, more modes implies more cusps. This is to be expected for several reasons. First of all, a higher RMS amplitude as well as more modes imply more energy in the string's vibrations. More energy means a higher average energy and favours highly relativistic points. On a more specific point of view, these high amplitudes and numerous modes imply large deviations from a straighter line, both for the physical string and for the curves ${\bf a'}$ and ${\bf -b'}$. This implies a wavier string, hence more crossing on the unit sphere.
\begin{figure}[t]
	\begin{center}
	\includegraphics[keepaspectratio, width=14cm]{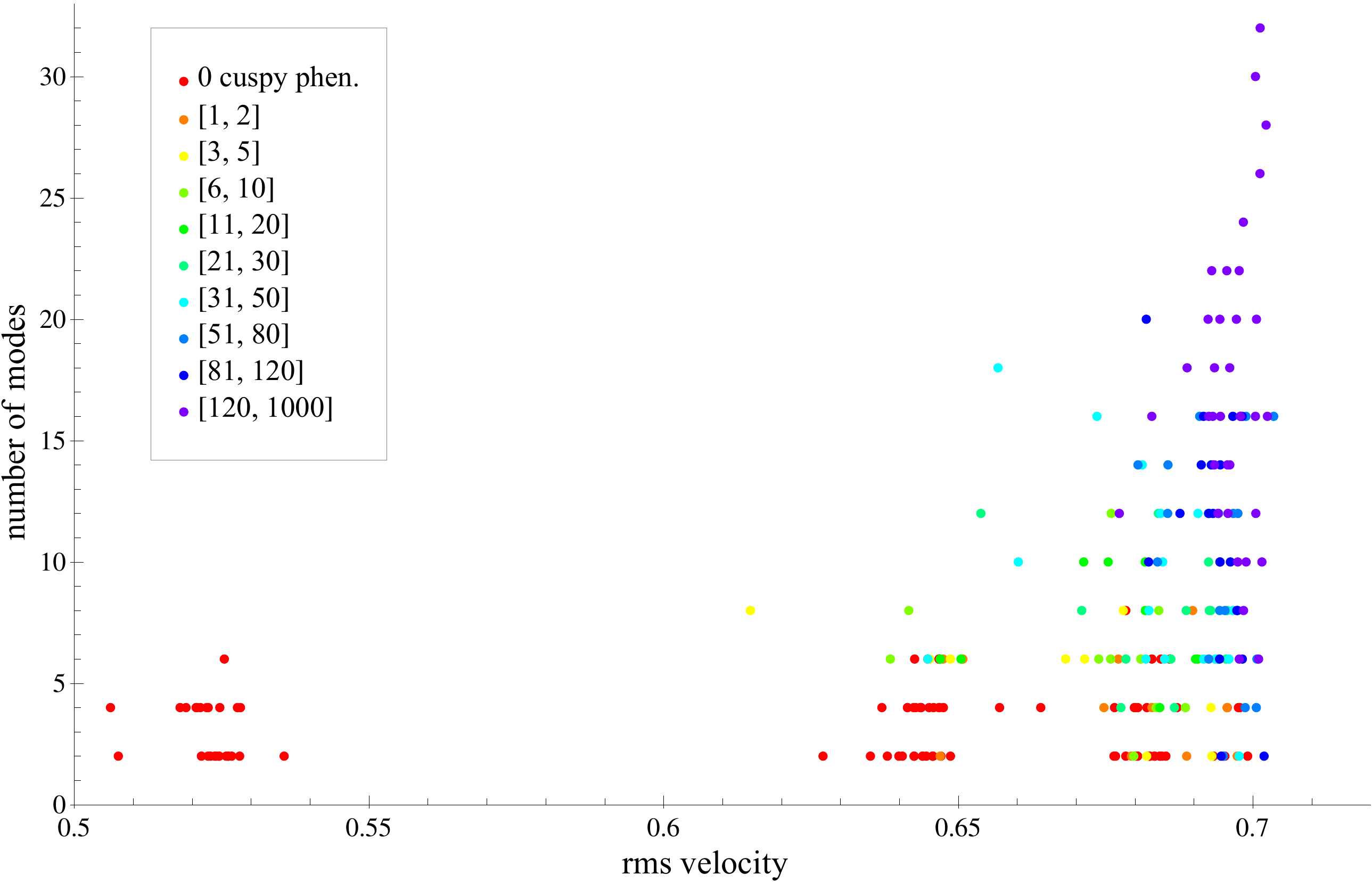}
	\caption{Number of $x$-modes versus the root mean square velocity of the string.\newline
		From red to purple, strings with $0$ to between $120$ and $1000$ cuspy events.$\qquad$}
	\label{RMSvelColCP}
	\end{center}
\vspace{-1em} \end{figure}
\begin{figure}[t]
	\begin{center}
	\includegraphics[keepaspectratio, width=14cm]{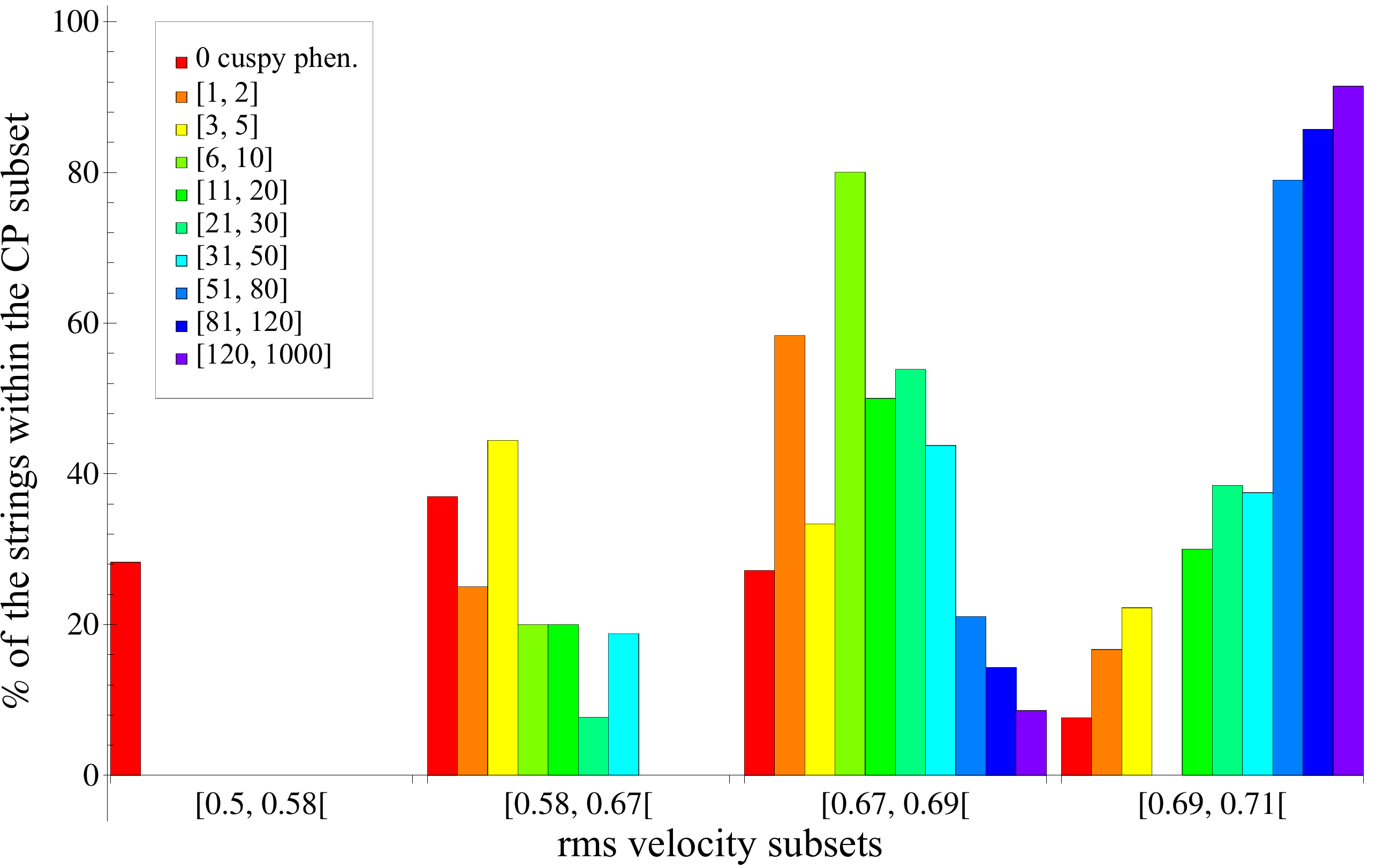}
	\caption{Bar chart of the percentage of the strings within a CP subset whose RMS velocity\newline
		lies in each interval. Same colour representation as previously.$\qquad$}
	\label{BarChtRmsVNbStrCP}
	\end{center}
\vspace{-1em} \end{figure}

One can then study the correlation with the RMS velocity of the string, which is related to what we just mentioned; we plot it on Fig.~\ref{RMSvelColCP} the number of $x$-modes versus the RMS velocity\footnote{We are here talking about the time-averaged root mean square velocity along the string.} of the string; again, a colour gradient is representing the strings grouped according to the number of cuspy events (from $0$ in red to above $120$ in purple). One can first notice that the RMS velocity reaches a maximum around $0.7$--$0.71$. This is due to the Virasoro and gauge conditions used on the finite string; indeed, it implies for the RMS velocity: $v_{\rm rms}^2 < \sfrac{1}{2} \;\;\Leftrightarrow\;\; v_{\rm rms} < \sqrt{0.5} \simeq 0.707$.

In addition to the previously studied correlation between the number of cuspy events and the number of $x$-modes, there is a strong dependence on the RMS velocity of the string, as expected. One can split the set of strings in four groups according to their RMS velocity: below $0.58$, between $0.58$ and $0.67$, between $0.67$ and $0.69$ and above $0.69$. While the first subset of string shows no cusps or pseudocusps, the last one contains almost all the strings with more than $120$ cuspy events and almost no string without any.

To be more explicit, for each subset of strings grouped according to the number of cuspy events, Fig.~\ref{BarChtRmsVNbStrCP} shows the percentage of strings in each interval of RMS velocity. One can indeed notice that in the highest interval (that is for RMS velocity above $0.69$) one only finds a few of the strings without cusps or pseudocusps (about $8\%$) but most of the strings with more than $50$ cuspy events ($80$ to $90\%$ of them). We also computed the average number of cuspy events in each of the four RMS velocity subsets and obtained:
\begin{align*}
	0. \, \pm \, 0. &\mbox{ cuspy phen. for strings whose RMS velocity is in } [0.50,\,0.58]\\
	4.3 \, \pm \, 1.5 &\hspace{1.5em} \mbox{"} \hspace{2em} \mbox{"} \hspace{2em} \mbox{"} \hspace{2em} \mbox{"} \hspace{2.5em} \mbox{"} \hspace{2em}
	\mbox{"} \hspace{3em} \mbox{"} \hspace{1.5em} \mbox{"} \hspace{0.7em} \mbox{"} \hspace{0.4em} [0.58,\,0.67]\\
	21 \, \pm \, 3.9 &\hspace{1.5em} \mbox{"} \hspace{2em} \mbox{"} \hspace{2em} \mbox{"} \hspace{2em} \mbox{"} \hspace{2.5em} \mbox{"} \hspace{2em}
	\mbox{"} \hspace{3em} \mbox{"} \hspace{1.5em} \mbox{"} \hspace{0.7em} \mbox{"} \hspace{0.4em} [0.67,\,0.69]\\
	130 \, \pm \, 16 &\hspace{1.5em} \mbox{"} \hspace{2em} \mbox{"} \hspace{2em} \mbox{"} \hspace{2em} \mbox{"} \hspace{2.5em} \mbox{"} \hspace{2em}
	\mbox{"} \hspace{3em} \mbox{"} \hspace{1.5em} \mbox{"} \hspace{0.7em} \mbox{"} \hspace{0.4em} [0.69,\,0.71]
\end{align*}
There is again an interesting correlation between the RMS velocity of the string, which is closely related to the energy of the string, and the number of cusps and pseudocusps.
\begin{figure*}[t]
	\begin{adjustwidth}{-2.5em}{-2.5em}
	\includegraphics[height=13.501cm, width=17cm]{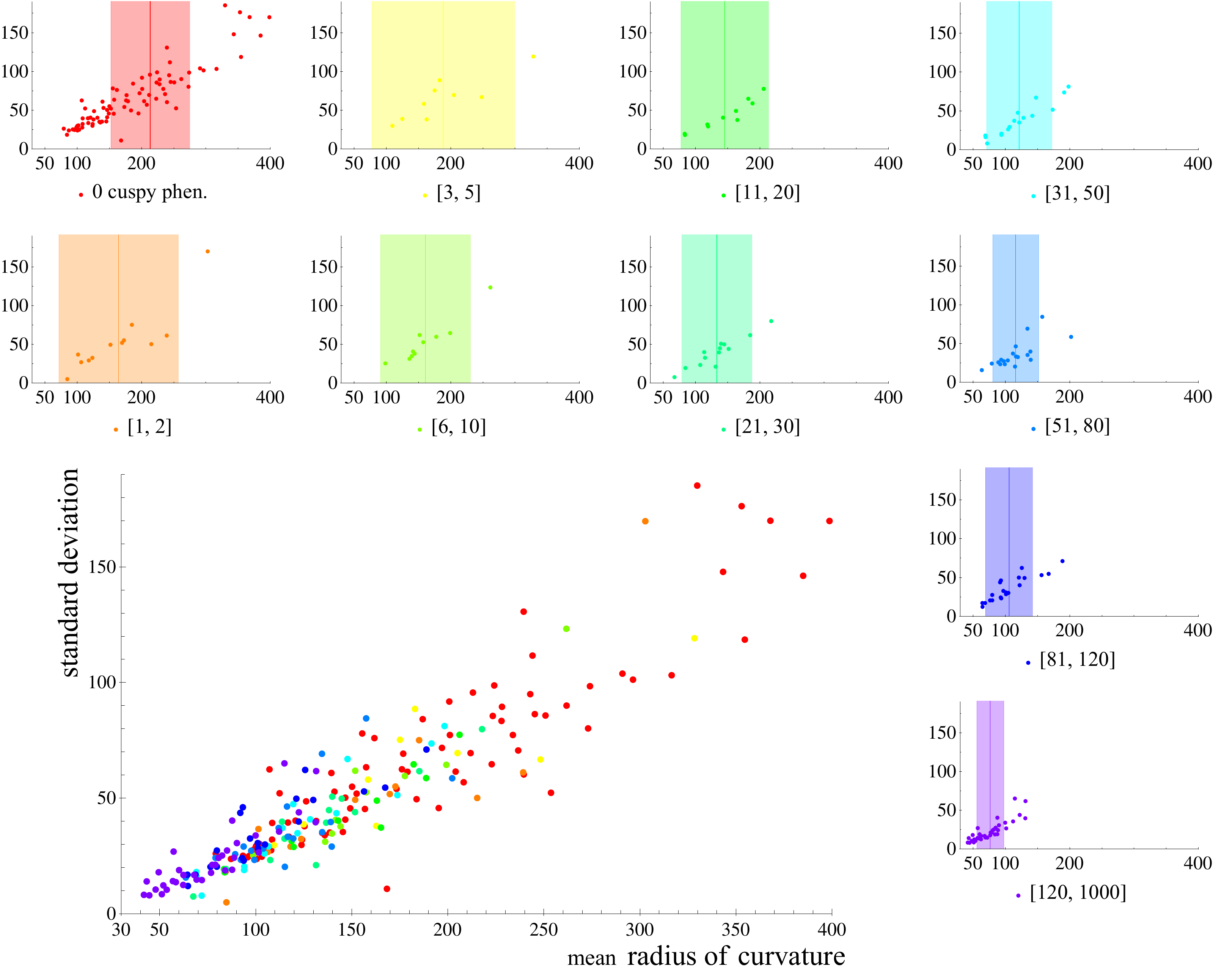}
	\caption{Radius of curvature and its standard deviation.\\
		Same colour representation. On each subgraph, the solid line marks the mean\\
		within the subset and the shaded area represents $5$ times the standard deviation.}
	\label{radcurvGrid}
	\end{adjustwidth}
\vspace{-0.7em} \end{figure*}

Finally, and in order to return to a previously mentioned concern, one might want to look at the correlation with the radius of curvature along the string.\footnote{We are here dealing with the radius of curvature averaged along the string. For clarity, in the following we call \emph{(mean) radius of curvature} the time average of the already space-averaged radius of curvature for each string separately; the \emph{standard deviation} of the radius of curvature is then the deviation during a period of time from this average. We thus end up with two figures per string.} Indeed, it can in turn be linked to the large-amplitude waves' characteristic length since it represents the average size of waves on the string; note though that it is several times larger than the characteristic length since it also takes into account the flat parts of the string between such waves.

With this in mind, we plot the standard deviation versus the (mean) radius of curvature for each string. We split up the set of strings according to the number of cuspy events and also draw the superposition of all the subgraphs. Figure~\ref{radcurvGrid} shows, from top to bottom and from left to right, the ten subgraphs along with the overall graph in the bottom left corner. For each separate subset has been computed the mean and the standard error\footnote{Here, the mean and the standard error are computed among the strings of a same subset on the (mean) radius of curvature, giving us two figures for each subset. Note that we define the standard error as $\frac{\sigma}{\sqrt{N_s}}$ where $\sigma$ is the standard deviation in the subset and $N_s$ is the number of strings in this subset.} of the radius of curvature, showing how it evolves with the number of cuspy events. They have been added via a solid line on the mean and a coloured shaded area around it encompassing $5$ times the standard deviation.

First of all, one can notice that the standard deviation grows almost linearly with the mean radius of curvature, albeit with some dispersion at large values. More interestingly, the radius of curvature is smaller for strings with many cusps: this shows again the foreseen correlation according to which a wavier string presents more cusps and pseudocusps. This can be seen from the overall graph, on which for instance points with a radius of curvature larger than $200$ have generally less than $5$ cuspy events, most of them having none. It can also be deduced from the subgraphs in Fig.~\ref{radcurvGrid}. More precisely, the mean of each subset is decreasing with the number of cuspy events: from $210$ for non-cuspy strings to $75$ for very cuspy ones. The standard deviation is also decreasing, apart from the less populated subsets (for instance, subsets of strings with $1$ to $5$ cusps and pseudocusps have larger standard deviation than the one for non-cuspy strings since the latter includes many more strings).

\subsection{Correlation with the parameters of the network} \label{sec:correlparam}

As mentioned previously, we are mainly interested in two networks' parameters: the interstring distance $\xi$ and the coherence length $\bar \xi$. We have defined $\Delta$ to be the distance between the junctions, hence it could be considered as the interstring distance (since it is the distance between two heavy strings) but physically, the ratio with the parameter length is more relevant. In our simulation, the end-to-end distance is fixed and the parameter length of the string plays a scaling r\^ole. Indeed, it turns the ratio $\sfrac{\Delta}{\sigma_m}$ into our length parameter since it gives the sum of the average vectors $\langle {\bf a'} \rangle_{\sigma}$ and $\langle {\bf b'} \rangle_{\sigma}$ --- which is along the $x$-axis --- in the unit sphere description. We can thus associate the interstring distance with this simulation's parameters ratio
\be
	\xi \sim \frac{\Delta}{\sigma_m}~.
\vspace{0.7em} \ee
In the case of a {\sl double} network consisting of both heavy and light strings, each one is associated with a set of parameters: $\xi_{light}$, ${\bar \xi}_{light}$ and $\xi_{heavy}$, ${\bar \xi}_{heavy}$. In agreement with the configuration we are studying, our analysis does not take into account the light string network's interstring distance $\xi_{light}$ but only the heavy one's via $\xi_{heavy} \sim \sfrac{\Delta}{\sigma_m}$.

The definition of the coherence length is more subtle for several reasons. First of all, our simulation do not input directly a typical length apart from the minimal wavelength of the vibrations on the string. Instead, random numbers are drawn to define the string's structure, implying that we need to compute afterwards the length scale. In addition, in our numerical approach, one may use different ways to define the characteristic size for waves and wiggles on the string and even different definitions of large-amplitude waves.

Still, let us explore some of the possibilities, starting with the usual definition \cite{Austin:1993rg,Vincent:1996rb} computing the correlation between two points along the string via
\be
{\bar \xi} \equiv \int^\infty_0 \!{\rm d} \sigma \;\langle a_x' (\sigma_1+t)\,.\, a_x' (\sigma_2+t) \rangle~.
\ee
But because our strings are by construction fully correlated, this definition is of no use. Indeed, defining the string's position with a Fourier decomposition implies that the whole string is correlated; mathematically, this result comes from the fact that the average of a sum of sines and cosines is $0$, resulting in ${\bar \xi} = \sigma_m$. Thus, we need to define our persistence length differently.

In the search for different formulations, one could think of the radius of curvature. This number defines for each string a condensed typical size of all the ripples on the string during the whole period. Unfortunately, it takes into account the flat parts of the string whose radius of curvature is obviously very large. This makes the strings' radius of curvature difficult to use in order to define a specific length scale but still allows us to notice some correlation: the number of cusps and pseudocusps grows with smaller radii of curvature. This means that the information about the large-amplitude waves is, at least partially, encoded in the radius of curvature even if we cannot simply access it.

Let us use what seems to be the simplest and most reliable way to define a scale for the large-amplitude waves on the string: the vibrations' frequency.  Indeed, the modes set up on the string at $t=0$ are stable and keep the same amplitude during the evolution. Even if they can be hidden at a specific time by other frequencies and not visible when looking at the string itself (or at its radius of curvature), they are characteristic of the way the string vibrates. Moreover, this parameter can be easily controlled from the input to the simulation and also evaluated once the string is drawn. The only remaining issue has to do with the number of the largest frequencies to be accounted for. Obviously, we could not only use the lowest frequency, that is the largest wavelength, because it would not take into account the waves on the string --- especially in our case where the largest wavelength is fixed and equal to twice the length of the string. We could use the highest frequency only and define the large-amplitude waves characteristic length directly according to the associated wavelength. This is not ideal though beauce there could be configurations where the highest frequency mode's amplitude is very small compared to that of the second highest frequency. This would indeed distort the data by increasing the highest frequency (compared to the physically relevant one), thus decreasing the interesting length scale. In general, this definition would also be too sensitive to the high frequency part of the Fourier decomposition and not enough to the whole spectrum.

One way to deal with this issue is to compute a length scale based on all the wavelengths $\lambda_k \equiv \sfrac{\sigma_m}{\,k}$, taking each one into account according to their rank $k$ and to the associated amplitude $A_k$.\footnote{Even if the amplitudes are drawn in a symmetric interval around $0$, one of them being actually null is statistically insignificant. This implies that the $k$-th wavelength is of the form $\sfrac{2\,\sigma_m}{k}$, recalling that the fundamental excitation has no nodes and thus has a wavelength equal to twice the string's length.} Different possibilites have been considered but what seemed to be the most accurate and the simplest one is to use the average wavelength ${\bar \lambda}$. One has to note first that in order to keep the velocity below $c=1$ at all time, one needs to choose amplitudes such that $A_k \sim \lambda_k$ (under the simplifying assumption that all modes carry roughly the same amount of energy). Keeping this in mind, looking at $\sum \sqrt{A_k^2 + \lambda_k^2}$ is equivalent to considering $\sum \lambda_k$.

Hence, we define the coherence length in terms of the mean wavelength ${\bar \lambda} \equiv 2\,\sigma_m \,H_{\bar n} \, / \,{\bar n}$, giving
\be
	\bar \xi \sim \frac{{\bar \lambda}}{4} = \frac{\sigma_m \, H_{\bar n}}{2\,{\bar n}} \simeq \frac{\sigma_m \left( \ln({\bar n}) + \gamma \right)}{2\,{\bar n}}~,
\ee
where ${\bar n}$ is the highest frequency mode on the string (and again not the parameter $n$ of the simulation) and $H_n = \sum^n_{k=1} \sfrac{1}{\,k}$ is the harmonic series. Recall $H_n \simeq \ln(n) + \gamma$ with $\gamma \simeq 0.577$ and that the difference $H_n - \ln(n) - \gamma$ is larger than $10\%$ of $H_n$ only for $n \le 3$, meaning that the approximation is sufficient for our estimation as soon as $n>3$. Finally, note that since the number of modes is quite low in our simulation (at most $16$ modes are taken into account), this cannot overlap with a definition of the wiggliness $\zeta$.

We have here estimated the two parameters of our strings' network in terms of two parameters of the simulation.\footnote{We used three parameters --- $\Delta$, $\sigma_m$ and $n_{\mbox{\scriptsize eff}}$ --- but in fact $\Delta$ is not a variable, leaving two actual parameters.} As foreseen, the parameter length of the string $\sigma_m$ plays an important r\^ole, both for defining the interstring distance and the coherence length. The number of modes seems like the most obvious and accurate way to define a large waves length scale.

\section{Conclusions}

Gravitational waves, even though they have yet to be observed, are at the center of attention. They are the next tool for cosmology and high energy astrophysics and should soon give us a stream of new data to analyse. Similarly, cosmic strings are thought to be unavoidable in most of the cosmic scenarii and should provide insight into the symmetry breaking they are remnants of or the theory to which they belong.

In this study, we have concentrated on a particular configuration made of a light string stretched between two junctions with heavy strings. It is important to note that even if we considered simplifying assumptions, the overall behaviour and the results should remain in more realistic configurations as long as the end points of the light string can be seen as fixed during a period of oscillation. We then looked at highly relativistic points since they are sources of high frequency bursts of gravity waves. Such cuspy events appear on a string when the left- and right-movers' velocities are temporarily equal (or approximately equal), making them reasonably easy to identify. We split them into two classes: the actual cusps, resulting from crossings of the two movers' velocity curves and hence reaching momentarily the speed of light $c=1$, and the so-called pseudocusps, resulting from a close approach between the two curves and hence reaching highly relativistic velocities, typically below $c=1$ by $10^{-3}$ to $10^{-6}$.

Since cuspy events emit large amounts of energy in the form of gravitational wave bursts, to estimate the signal that could be detected in the neighbourhood of the Earth by ground- and space-based detectors, one needs to know how frequently they occur. We have here aimed to quantify this and analyse it in terms of the parameters characterising the string configuration, as well as the string network through the usual network parameters $\xi$ and $\bar \xi$ (but not~$\zeta$).

Our analytical approach allowed us to identify the symmetries of the problem. Indeed, because of the boundary conditions, the string moves (almost) always periodically. In addition, on the unit sphere, the left- and right-movers' velocities are symmetric with respect to the axis parallel to the heavy strings. This simplifies the problem enough to evaluate the frequency of cusps and pseudocusps on the string with respect to a few parameters.

We found that cusps should be frequent for strings satisfying (see Eq.~(\ref{eq:criterion})):
\be
	\langle a_x' a_x' \rangle_\sigma \gtrsim \frac{1+\alpha}{\alpha} \left( \frac{|\Delta|}{\sigma_{m}} \right)^2~, \nonumber
\ee
where ${\bf a}$ is the left-mover on the string, $|\Delta|$ the end-to-end vector's norm and $x$ its direction (the subscript $x$ thus referring to the projection on the $x$-axis), $\sigma_m$ the parameter length of the string and $\alpha$ a parameter we subsequently estimated around $\alpha \simeq 4.1$. It is important to notice that such cuspy strings should present many important waves.

We then used a simulation to get a statistically important number of strings within a range of parameters, in order to check this behaviour. The set of $237$ strings we obtained presents $8719$ cusps and $4659$ pseudocusps, i.e. there are slightly less than half the number of cusps --- as roughly expected. We analysed the occurrence of cuspy events with respect to several other features, confirming our analytical work and the general behaviour of such strings.

In particular, we first checked that our characterisation of pseudocusps from the minimal angle between the two curves on the unit sphere is relevant. For instance, the velocity we obtained from this description is very close to the one obtained directly from the simulation (within grid and computational inaccuracies). In addition, the presence of cusps and pseudocusps increases according to the inequality Eq.~(\ref{eq:criterion}), giving us an accurate tool to discriminate between cuspy and non-cuspy strings. More importantly, it also depends on the number and amplitude of the vibration modes in the $x$-direction; this confirms more directly the fact that the wavier a string is, the more cuspy events it presents.

We also analysed the influence of the RMS velocity on the string: as one could expect, the more energy there is on the string, the more cusps appear. This is consistent with the fact that more vibrating modes imply more cusps, since both indicate more energy. Finally, we found the radius of curvature along the string is also correlated to the number of cusps and pseudocusps, favouring again the mentioned behaviour (a smaller radius of curvature is equivalent to more waves, which are in turn linked to more cusps).

Expressing the usual network parameters in terms of our simulation's parameters, we refined the link between the numerical description and the way cosmic strings networks are traditionally pictured. This should allow future work, whether on gravitational waves or on interacting evolution of the network, to assess, use and further continue this work.

\begin{acknowledgments}
The work of W.~N. was partly supported by the Marie Curie Fellowship program of the EU.
\end{acknowledgments}

\appendix 
\section{Generalised strings' configuration}
We here extend our initial strings' configuration detailed in Section~\ref{sec:setup}, in order to show that the quasi-periodicity of the movement of the light string is indeed generic.

\subsection{Coplanar heavy strings with various angles}
\begin{figure*}[t]
	\begin{center}
	\includegraphics[height=6.38cm, keepaspectratio]{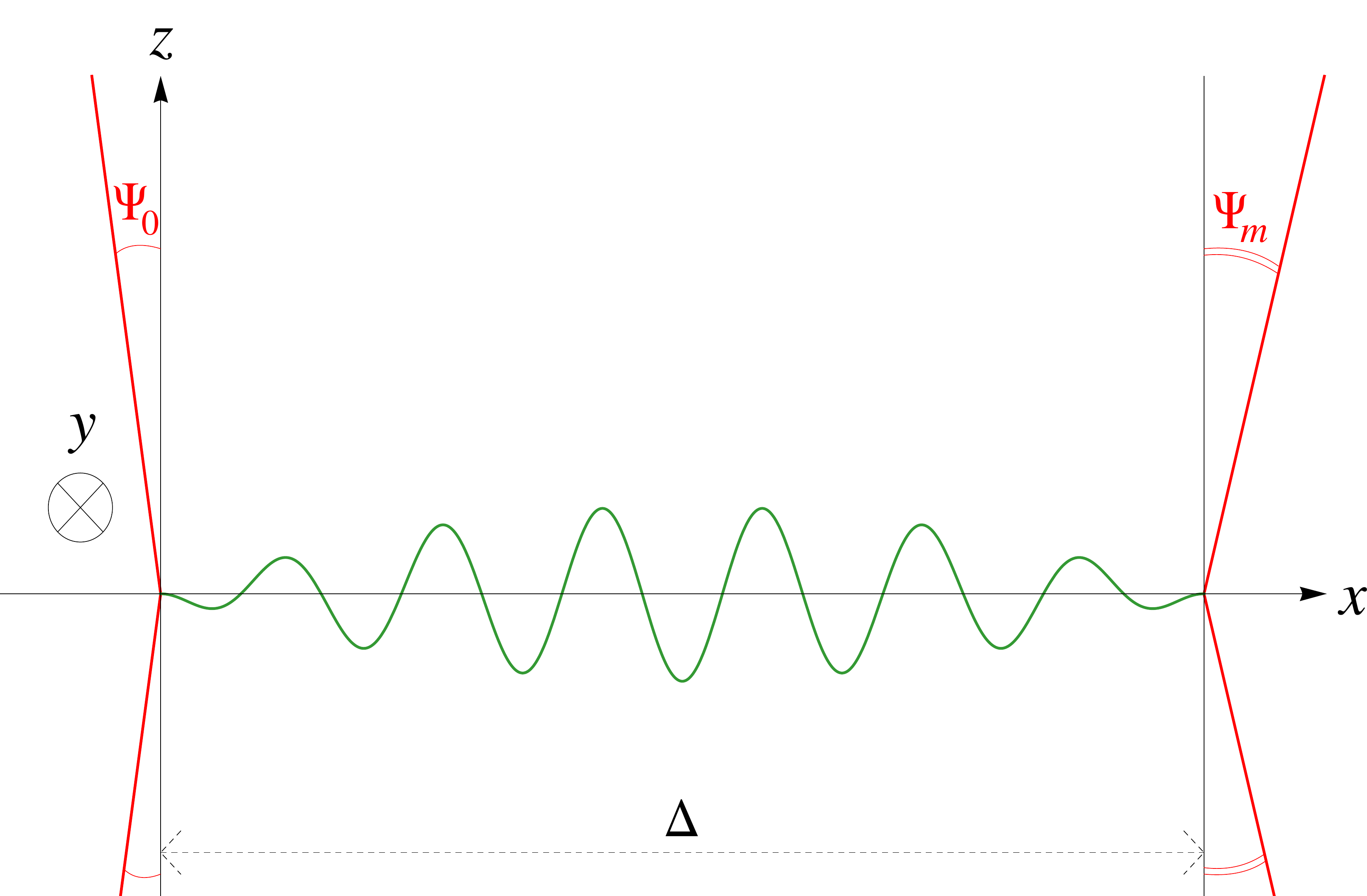}
	\caption{A light string stretched between two junctions with heavy strings.\\
		Here the upper-half plane is symmetric to the lower-half plane and each heavy string\\
		forms a different angle with the $z$-axis. The heavy strings are coplanar.}
	\label{StringLayout_Ang}
	\end{center}
\vspace{-1em}\end{figure*}

In this section, we choose different angles at the two junctions and denote $\Psi_0$ (respectively $\Psi_m$) the angle between the $z$-axis and the heavy string at the $\sigma=0$ (respectively $\sigma=\sigma_m$) junction. In addition, by setting the upper half-plane to be the symmetric of the lower half-plane, one forms a $(\pi - 2 \Psi_0)$ (respectively $(\pi - 2 \Psi_m)$) angle along the heavy string. Note that here, the two heavy strings remain coplanar and orthogonal to the $y$-axis, as shown on Fig.~\ref{StringLayout_Ang}.

One can then define $S_0 = \mathrm{sign}( x_z'(0,t) )$ and $S_m = \mathrm{sign}( x_z'(\sigma_m,t) )$ the signs of the \mbox{$z$-component} of the light string's velocity at each end. These both take the value $+1$ or $-1$ depending on whether we consider the $z<0$ or $z>0$ half-plane, respectively. They allow us to write in a compact way all the boundary conditions coming from Eqs.~(\ref{condattheborder}), giving
\begin{subequations}
	\begin{align}
	\dot{x}_y \left( t,0\right) &= 0~, \\
	\dot{x}_x \left( t,0\right) - S_0 \tan (\Psi_0) \; \dot{x}_z\left( t,0\right) &=0~, \\
	S_0 \tan (\Psi_0) \; x'_x \left( t,0\right) + x'_z\left( t,0\right) &= 0~,
	\end{align}
and\vspace{-1.2em}
	\begin{align}
	\dot{x}_y \left( t,\sigma_{m}\right) &= 0~, \\
	\dot{x}_x \left( t,\sigma_{m}\right) + S_m \tan(\Psi_m) \; \dot{x}_z\left( t,\sigma_{m}\right) &= 0~, \\
	S_m \tan(\Psi_m) \; x'_x \left( t,\sigma_{m}\right) - x'_z\left( t,\sigma_{m}\right) &= 0~,
	\end{align}
\end{subequations}
leading to the system of equations
\begin{subequations} \label{AandBx6_ang}
	\begin{align}
	a_y' \left( t \right) &= b_y' \left( -t \right)~, \label{eq:1_ang}\\
	\left[\, a_z' \left( t \right) - b_z' \left( -t \right) \:\!\right]\, S_0 \tan \Psi_0 &= a_x' \left( t \right) - b_x' \left( -t \right)~, \label{eq:2_ang}\\
	a_z' \left( t \right) + b_z' \left( -t \right) &= - \left[\, a_x' \left( t \right) + b_x' \left( -t \right) \:\!\right]\, S_0 \tan \Psi_0 ~, \label{eq:3_ang}
	\end{align}
and\vspace{-1.2em}
	\begin{align}
	a_y' \left( 2 \sigma_{m}+t \right) &= b_y' \left( -t \right)~, \label{eq:4_ang}\\
	\left[\, a_z' \left( 2 \sigma_{m}+t \right) - b_z' \left( -t \right) \:\!\right]\, S_m \tan \Psi_m &= - a_x' \left( 2 \sigma_{m}+t \right) + b_x' \left( -t \right)~, \label{eq:5_ang}\\
	a_z' \left( 2 \sigma_{m}+t \right) + b_z' \left( -t \right) &= \left[\, a_x' \left( 2 \sigma_{m}+t \right) + b_x' \left( -t \right) \:\!\right]\, S_m \tan \Psi_m~, \label{eq:6_ang}
	\end{align}
\end{subequations}
replacing Eqs.~(\ref{AandBx6}). Manipulating Eqs.~(\ref{eq:2_ang}) and (\ref{eq:3_ang}) allows us to express $a_x' (t)$ and $b_x' (-t)$ in terms of $a_z' (t)$, $b_z' (-t)$ and polynomials of $(S_0 \tan \Psi_0)$, and thus $a_x' (2\sigma_m+t)$ after a shift $t \rightarrow 2\sigma_m+t$. Replacing in Eqs.~(\ref{eq:5_ang}) and (\ref{eq:6_ang}), one gets two equations involving $a_z' (2\sigma_m+t)$, $a_z' (t)$, $b_z' (-2\sigma_m-t)$ and $b_z' (-t)$, and combination of $(S_0 \tan \Psi_0)$ and $(S_m \tan \Psi_m)$.

Shifting the variable $t \rightarrow 2\sigma_m+t$, one gets four equations involving six variables: $a_z' (4\sigma_m+t)$, $a_z' (2\sigma_m+t)$, $a_z' (t)$, $b_z' (-4\sigma_m-t)$, $b_z' (-2\sigma_m-t)$ and $b_z' (-t)$. One can then use three of them to eliminate the three $b_z'$ variables --- namely $b_z' (-4\sigma_m-t)$, $b_z' (-2\sigma_m-t)$ and $b_z' (-t)$ --- to obtain an expression similar to Eq.~(\ref{eq:diff1})
\beq
	a_z' \left( t \right) = - {\cal R} a_z' \left( -2\sigma_{m}+t \right) - a_z' \left( -4\sigma_{m}+t \right)~,
\eeq
where
\be
	{\cal R} \equiv -2 \cos \left( 2 S_0 \Psi_0 + 2 S_m \Psi_m \right)~,
\ee
and similarly for $a_x'$.

This expression is very similar to the one we obtained in the initial setting, which we can retrieve by setting $S_m = 1 = S_0$ and $\Psi_0 = \Psi_m$. In addition, this equation also reveals that the functions $a_x'$ and $a_z'$ are periodic for a dense subset of angles, otherwise quasi-periodic; one simply needs to replace $2 \Psi$ by $\Psi_0 \pm \Psi_m$. This justifies our initial simpler choice.

\subsection{Non-coplanar heavy strings}
\begin{figure*}[t]
	\begin{adjustwidth}{-5em}{-5em}
	\begin{center}
	\begin{subfigure}{10cm}
	\includegraphics*[height=7.38cm, keepaspectratio, trim= 130 20 28 2, clip]{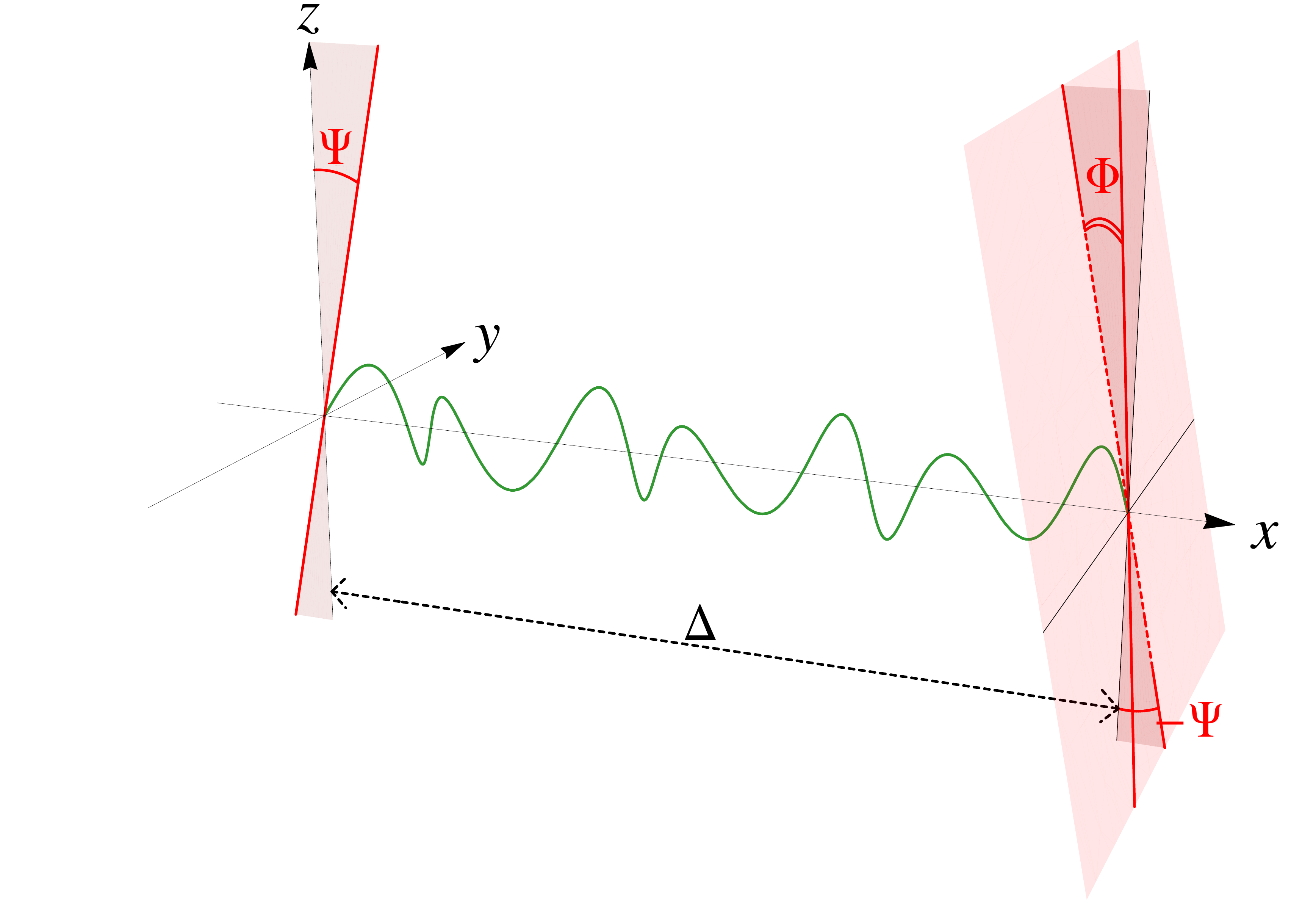}
        \setlength{\abovecaptionskip}{-5pt}
	\caption{front view}
	\label{StringLayout_NonCopl_Ft}
	\end{subfigure}
	\begin{subfigure}{7cm}
	\includegraphics*[height=6.38cm, keepaspectratio, trim= 190 30 60 2, clip]{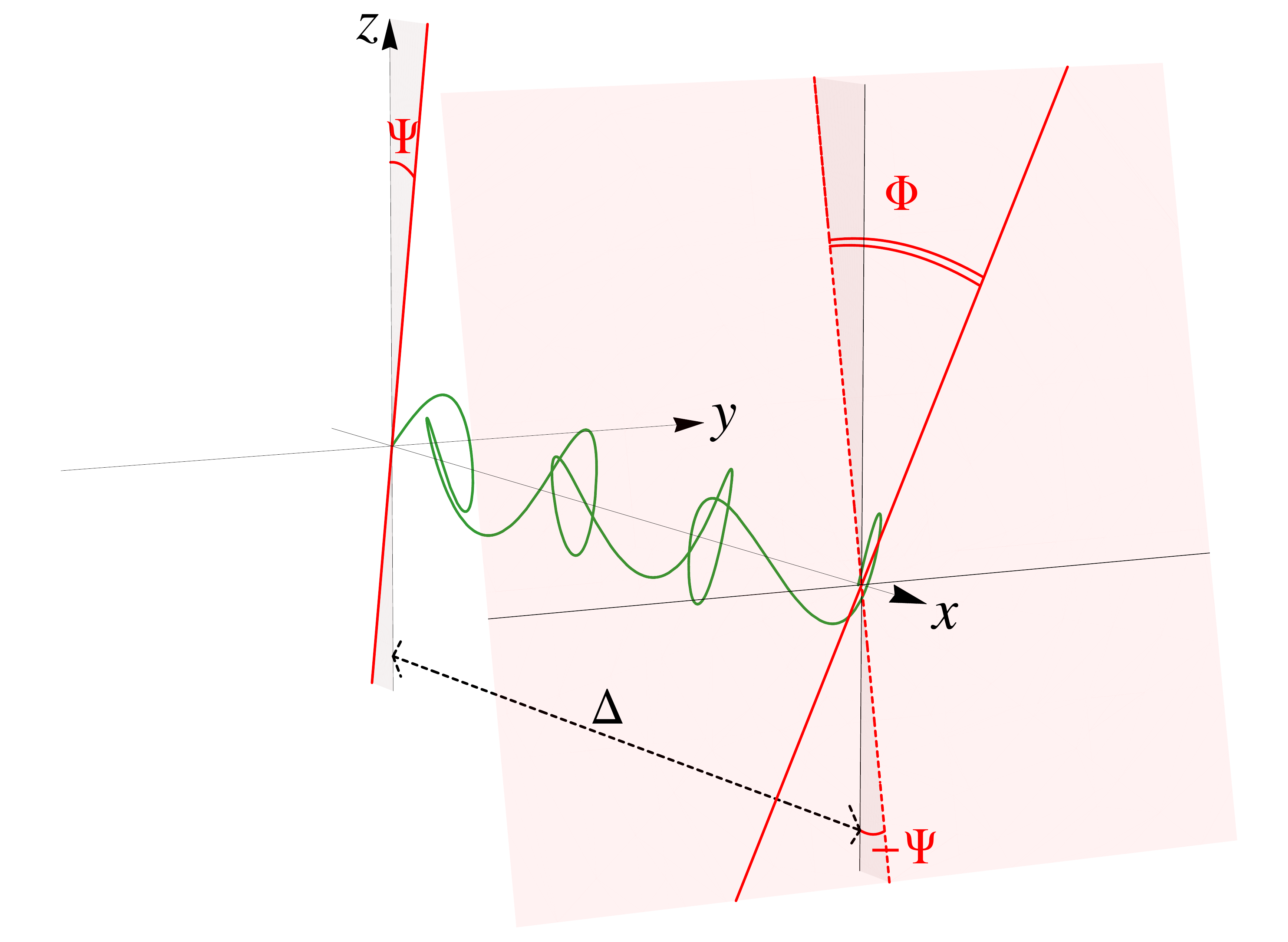}
        \setlength{\abovecaptionskip}{15pt}
	\caption{side view}
	\label{StringLayout_NonCopl_Sd}
	\end{subfigure}
	\caption{A light string stretched between two junctions with heavy strings.\\
		Here the $\sigma=\sigma_m$ end string has been rotated in the plane containing the $y$-axis\\
		by an angle $\Phi$. The two heavy strings are no longer coplanar.}
	\label{StringLayout_NonCopl}
	\end{center}
	\end{adjustwidth}
\vspace{-1em} \end{figure*}

In this section, we choose to modify the initial configuration by rotating the $\sigma_m$-end string in the plane containing the $y$-axis, as shown on Fig.~\ref{StringLayout_NonCopl}.
In other words, one rotates the string around the axis which is perpendicular both to the initial position of the string and to the $y$-axis, that is the axis directed by the vector $(\cos \Psi, \,0\,, \,\sin \Psi)$.

This rotation generates a coupling between $a'_y$ and the other components of $\bf{a'}$, namely $a'_x$ and $a'_z$, contrarily to previous cases. Indeed, the boundary conditions at $\sigma=0$ remain the same while the ones at $\sigma=\sigma_m$ become
\begin{subequations}
	\begin{align}
	-\sin \Psi \,\sin \Phi \left( a_x' \left( 2 \sigma_{m}+t \right) - b_x' \left( -t \right) \right) + \cos \Phi \left( a_y' \left( 2 \sigma_{m}+t \right) - b_y' \left( -t \right) \right) \qquad&\nonumber\\
		+ \cos \Psi \,\sin \Phi \left( a_z' \left( 2 \sigma_{m}+t \right) - b_z' \left( -t \right) \right) &= 0~, \label{eq:4_ncpl}\\
	\left( a_x' \left( 2 \sigma_{m}+t \right) - b_x' \left( -t \right) \right) + \tan \Psi \left( a_z' \left( 2 \sigma_{m}+t \right) - b_z' \left( -t \right) \right) &= 0~, \label{eq:5_ncpl}\\
	\sin \Psi \,\cos \Phi \left( a_x' \left( 2 \sigma_{m}+t \right) + b_x' \left( -t \right) \right) + \sin \Phi \left( a_y' \left( 2 \sigma_{m}+t \right) + b_y' \left( -t \right) \right) \qquad&\nonumber\\
		- \cos \Psi \,\cos \Phi \left( a_z' \left( 2 \sigma_{m}+t \right) + b_z' \left( -t \right) \right) &= 0~, \label{eq:6_ncpl}
	\end{align}
\end{subequations}
replacing Eqs.~(\ref{eq:4}) to (\ref{eq:6}). These are significantly more complicated than previously and imply that one needs to manipulate more equations to obtain a relationship similar to Eq.~(\ref{eq:diff1}). In the end, this coupling generates a 3$^{\mathrm{rd}}$ order equation for $a'_x$ and $a'_z$ instead of the 2$^{\mathrm{nd}}$ order one that is Eq.~(\ref{eq:diff1}).

We believe that the conclusion on the periodicity, obtained in the previous string configurations, is still valid in this general setup, basically since the energy density per unit length remains constant (no emission has been incorporated). Indeed, the energy being constant implies that any damping or amplification in one of the components of the signal along the string is linked to some compensation somewhere else in the system.

In the previous situations, if say the energy of the $y$-component was null at the beginning, it remained that way; similarly, the energy loss in say the $x$-component was balanced by the gain in the $z$-component. In our non-coplanar situation, one needs to take into account all three components in a very entangled and more complex way. This suggests that a loss of energy in say the $z$-component is going to be balanced by an amplification in say the $y$-component. Indeed, at the $\sigma=\sigma_m$ junction, this kind of transfer can happen since all three modes are coupled. In addition, it is believed that the damping in the $z$-direction could be seen as a source term in the $x$- and $y$-directions, linked to a general conservation of energy density and implying a globally periodic movement.

More precisely, the $3^\mathrm{rd}$ order equation is of the form
\be
	a_{n+3} - {\cal \bar R} \,a_{n+2} + {\cal \bar R} \,a_{n+1} - a_n = 0
\ee
where ${\cal \bar R}$ depends solely on the angles; it gives solutions of the form
\be
	a_n = A \,e^n + e^{un} \left( B \cos vn + C \sin vn \right)
\ee
where $A$, $B$ and $C$ are constants depending on the initial conditions (i.e. on $a_0$, $a_1$ and $a_2$) and $u$ and $v$ depend directly and solely on ${\cal \bar R}$. Taking $A$ to be non-zero gives unphysical solutions since one needs to keep in mind that $a'_y = \pm \sqrt{1 - (a'_x)^2 - (a'_z)^2}$.\footnote{Indeed, recall ${\bf a'}^2 = 1 = (a'_x)^2 + (a'_y)^2 + (a'_z)^2$.} One would get large values for $a'_x$ and $a'_z$ as $n$ grows, giving a negative value for $(1 - (a'_x)^2 - (a'_z)^2)$. Similarly, one cannot understand physically the exponential prefactor $e^{un}$ unless there is a mechanism to either suppress this factor or reverse it after some time. Indeed, let us divide this in three cases: if $u$ is null, one obtains a periodic motion; if $u>0$, we find ourselves in the case described previously, that is unphysical complex values for $a'_y$; finally, if $u<0$, one would have a situation where $a'_z = 0 = a'_x$ and all the energy lies in $a'_y$, which is unrealistic as well. A mechanism suppressing or reversing this prefactor would imply a balance between each component through time, which again makes sense physically.

Generally, it is believed that the rotation of the $\sigma=\sigma_m$ string should not change the global understanding of the movement of the light string, meaning that what was considered as consistent in the coplanar case should remain valid here.

\section{Snapshots of the simulation} \label{sec:snapshots}

\begin{figure*}[t]
	\begin{center}
	\includegraphics[height=18cm, keepaspectratio]{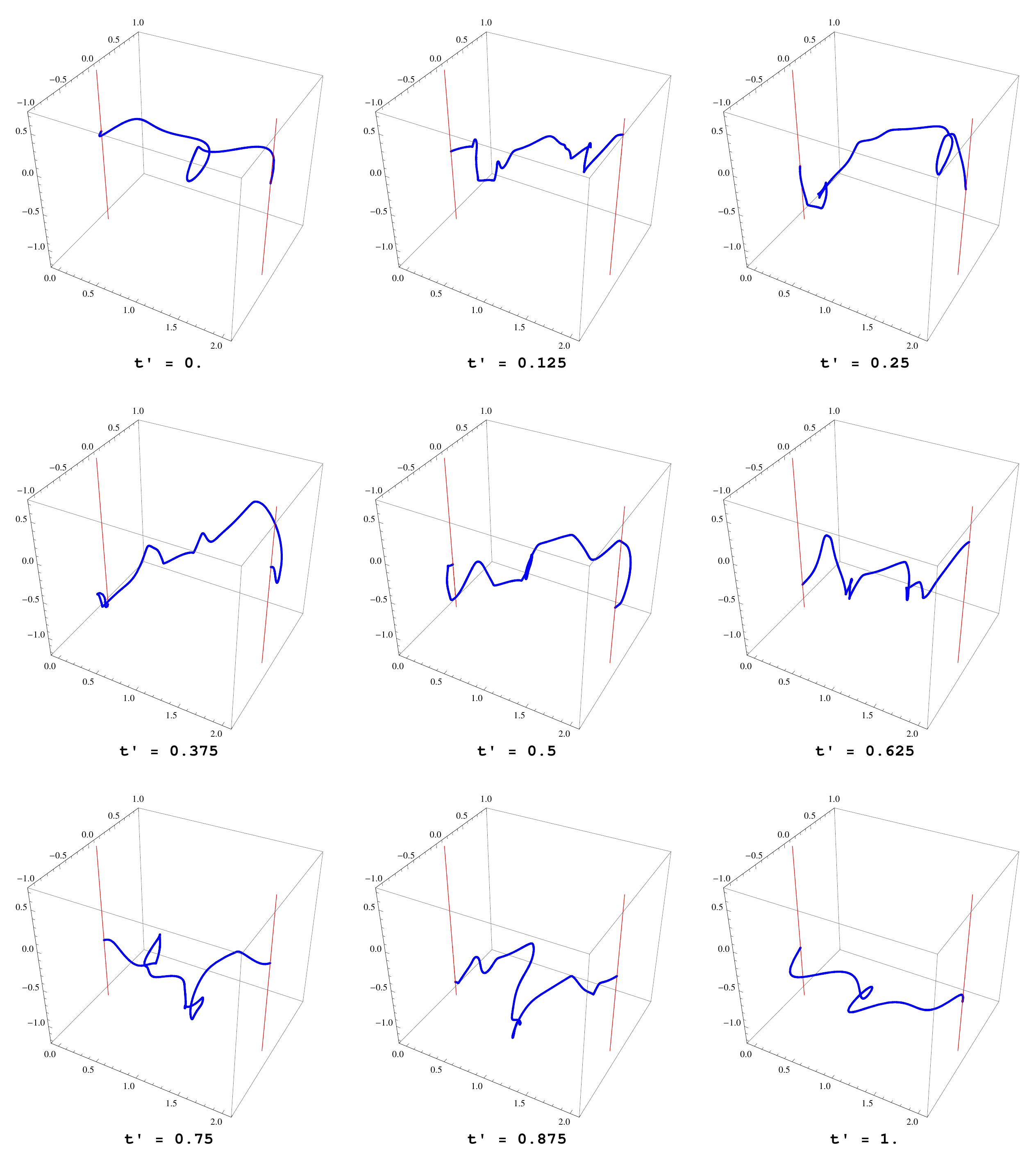}
	\caption{A simulated light string (in blue) stretched between two junctions\\
		with fixed heavy strings (in red). $t' = \sfrac{t}{\sigma_m}$ is the rescaled time.\\
		$\zeta \sim \sfrac{\Delta}{\sigma_m} = 0.25$ and $\bar \zeta \sim 2$.}
	\label{fig:snapshots}
	\end{center}
\vspace{-1em}\end{figure*}

We present on Fig.~\ref{fig:snapshots} some snapshots of a string simulated using our code. The chosen parameters here are such that $\zeta \sim \sfrac{\Delta}{\sigma_m} = 0.25$ and $\bar \zeta \sim 2$ since $4$ modes have been implemented on the string. Finally, we use here a rescaled time $t' \equiv \sfrac{t}{\sigma_m}$, meaning that $t' = 1$ after a half of the period. Note though that using symmetries, one can deduce how the string is behaving in the second half of the period from the string's position during the first half. Finally, note that $\Psi = 0$.



\end{document}